\documentclass[12pt]{article}

\usepackage{amsmath,amssymb,latexsym}
\usepackage[dvips]{graphicx}

\newcommand{\ol}{\overline}

\newcommand{\ra}{\rightarrow}

\newcommand{\cA}{{\cal A}}
\newcommand{\cC}{{\cal C}}
\newcommand{\cD}{{\cal D}}
\newcommand{\cF}{{\cal F}}
\newcommand{\cG}{{\cal G}}
\newcommand{\cO}{{\cal O}}
\newcommand{\cP}{{\cal P}}
\newcommand{\cR}{{\cal R}}
\newcommand{\cW}{{\cal W}}
\newcommand{\cY}{{\cal Y}}

\newcommand\p\partial
\DeclareMathOperator\Tr{\mathrm{Tr}}
\newcommand\be{\begin{equation}}
\newcommand\ee{\end{equation}}

\newcommand{\wt}{\widetilde}

\newcommand{\del}{\partial}
\newcommand{\nn}{\nonumber}

\newcommand{\half}{\frac{1}{2}}

\newcommand{\VEV}[1]{\left\langle #1\right\rangle}

\usepackage{xcolor}

\makeatletter
\@addtoreset{equation}{section}
\makeatother

\newcommand\req[1]{(\ref{#1})}

\begin{document}

\begin{flushright}
\hfill{YITP-15-103}\\
\hfill{IPMU15-0146}
\end{flushright}
\begin{center}
\vspace{2ex}
{\Large {\bf 
Holographic Chern-Simons Defects
}}

\vspace*{5mm}
{\sc Mitsutoshi Fujita}$^{a,b}$\footnote{e-mail:
 {\tt mitsutoshi.fujita@uky.edu}},
{\sc Charles M. Melby-Thompson}$^{c,d}$\footnote{e-mail:
 {\tt charlesmelby@fudan.edu.cn}},
{\sc Ren\'e Meyer}$^{e,d}$\footnote{e-mail:
 {\tt rene.meyer@stonybrook.edu}},
\\~~and~
{\sc Shigeki Sugimoto}$^{b,d}$\footnote{e-mail:
 {\tt sugimoto@yukawa.kyoto-u.ac.jp}}

\vspace*{4mm} 

\hspace{-0.5cm}
{\it {$^{a}$
Department of Physics and Astronomy, University of Kentucky, Lexington,
 KY 40506, USA}}\\ 
{\it {$^{b}$
Yukawa Institute for Theoretical Physics, Kyoto University,
Kyoto 606-8502, Japan
}}\\
{\it {$^{c}$
Department of Physics, Fudan University, 220 Handan Road, 200433
 Shanghai, China}}\\ 
{\it {$^{d}$ 
Kavli Institute for the Physics and Mathematics of the Universe (WPI),\\
 The University of Tokyo Institutes for Advanced Study (UTIAS),\\
 The University of Tokyo, Kashiwanoha, Kashiwa, 277-8583, Japan}}\\ 
{\it {$^{e}$
Department of Physics and Astronomy, Stony Brook University,
Stony Brook, New York 11794-3800, USA}}\\

\end{center}

\vspace*{.3cm}
\begin{center}
{\bf Abstract}
\end{center}

We study SU$(N)$ Yang-Mills-Chern-Simons theory in the presence of
defects that shift the Chern-Simons level from a holographic point of
view by embedding the system in string theory.  
The model is a D3-D7 system in Type IIB string theory, whose gravity dual is
given by the AdS soliton background with probe D7 branes attaching to
the AdS boundary along the defects. 
We holographically renormalize the free energy of the defect system with
sources, from which we obtain the correlation functions for certain
operators naturally associated to these defects. 
We find interesting phase transitions when the separation of the defects
as well as the temperature are varied.
We also discuss some implications for the Fractional Quantum Hall Effect
and for 2-dimensional QCD.

\newpage

\tableofcontents

\section{Introduction}

Pure gauge theory in three dimensions has some distinguishing features 
when compared to its more familiar 4-dimensional cousin, which arise 
from two important differences:  
(1) the Yang-Mills (YM) coupling is a dimensionful quantity and 
determines the scale of confinement; and
(2) in three dimensions it is possible to include a Chern-Simons (CS) 
term,
inducing a gauge-invariant ``topological mass''~\cite{Deser:1981wh}.

Pure Chern-Simons theory in particular has a wide range of interesting
properties, and has seen numerous applications. 
Witten \cite{Witten:1988hf} showed that the expectation values of 
Wilson loop operators in $SU(2)$ CS theory reproduce the Jones 
polynomial knot invariants, 
leading to extensive development of its applications to knot theory.
CS theory also has important applications in condensed matter
theory, the most important perhaps being its use as
an effective theory for the Fractional Quantum Hall Effect (FQHE). 
For example,
$U(1)$ CS-theory at level $m$ gives the low energy effective theory of 
the $m^{\rm th}$ Laughlin state, which realizes the
FQHE with filling fraction $\nu=1/m$ \cite{Zhang:1988wy,LeeZhang} (for a
review see, \emph{e.g.,} \cite{Zhang:1992eu}). 
Another interesting aspect of CS theory is level-rank duality, 
the equivalence between the $U(N)$ CS theory at level%
\footnote{In this paper, the CS level always denotes the bare value in the 
YM regularization.} 
$k$ and the $U(k)$ CS theory at level $N$
\cite{Naculich:1990pa,Camperi:1990dk,Mlawer:1990uv,
Naculich:2007nc}. \footnote{More precisely, as reviewed in
\cite{Aharony:2015mjs}, there are several flavors of level-rank duality,
the ones relevant to the unitary group being the
$SU(N)_k \leftrightarrow U(k)_{-N,-N}$ and 
$U(N)_{k,k+N}\leftrightarrow U(k)_{-N,-k-N}$ dualities.
(Here the first and second subscripts of $U(m)$ denote the levels of 
the $SU(m)$ and the $U(1)$ components, respectively.}
This level-rank duality is related to Seiberg-like duality
in 3-dimensional supersymmetric gauge theories
\cite{Aharony:2008gk,Giveon:2008zn,Niarchos:2008jb,Benini:2011mf,
Aharony:2013dha,Aharony:2014uya}.
More recently, it has been generalized to non-supersymmetric theories
with matter in the fundamental representation
\cite{Giombi:2011kc,Aharony:2012nh,GurAri:2012is,Aharony:2012ns,
Jain:2013py,Jain:2013gza,Aharony:2015mjs},
providing rare examples of dualities in dynamical theories that can 
be established explicitly by exact calculations.

In this paper, we study the properties of 2-dimensional defects
(domain walls) separating two phases at different CS levels 
in $SU(N)$ YM-CS theory using string theory and
holography.\footnote{The system we consider is not exactly YM-CS theory,
but contains extra massive matter. See section \ref{braneconfig} for
details.}
While the CS term is not gauge-invariant in the presence of
such level changing defects, the full quantum system is rendered 
consistent by the Callan-Harvey anomaly inflow mechanism
\cite{Callan:1984sa}: 
the tree level gauge variation of the action is canceled 
by the chiral anomaly of chiral fermions that live on the defect.

There are several motivations for introducing such defects.
One is to see how the defects behave under 
level-rank duality. 
Assuming that level-rank duality works locally,
it predicts that the rank of the gauge group in the dual description 
jumps at the defect.%
%
\footnote{%
The recent paper \cite{Armoni:2015jsa} studied similar defects in 
supersymmetric CS theory, together with their behavior under level-rank 
duality, in terms of intersecting brane models and their brane moves.}
We will give a geometric understanding of this
phenomenon in terms of the brane configuration in the holographic
model. 
A second motivation arises
if we compactify one of the spatial directions to a circle, 
and introduce a defect anti-defect pair separated along the circle; 
the system then flows to 2-dimensional QCD at low energies. 
Two-dimensional
QCD in the large $N$ limit is known to be solvable \cite{'tHooft:1974hx}
and has rich structures, such as confinement and chiral symmetry breaking,
similar to 4-dimensional QCD. We will discuss some interesting relations
between 3-dimensional YM-CS theory and 2-dimensional QCD, which might
shed some new light on the QCD physics.
These defects are of interest in condensed matter physics as
well, so it is of value to study them 
in the holographic context. 
As we will see, these defects generalize edges in the FQHE, and like 
FQH state edges, have gapless chiral excitations localized on them.

A string theory realization of the YM-CS system (without defects) was
proposed in~\cite{Fujita:2009kw}, where YM-CS dynamics is realized as
the infrared behavior of a D3/D7 system. The 3-dimensional $SU(N)$ gauge
theory with level $(-k)$ CS term%
\footnote{We take this unusual sign convention for the level because it
turns out to be convenient when discussing the holographic dual. 
This is related to the sign change under level-rank duality of Chern-Simons 
theory: the $U(k)_{N,N}$ theory is dual to $SU(N)_{-k}$.}
is constructed by putting $N$ D3 branes 
compactified on an $S^1$ with SUSY-breaking boundary conditions 
and $k$ units of RR 1-form flux.
The gravity dual is obtained by taking the near horizon limit of the
background corresponding to the $N$ D3 branes (the AdS${}_{5}$ soliton) in
the presence of $k$ probe D7 branes wrapped on $S^5$. 
One nice feature of this construction is that its IR behavior explains
the level-rank duality of CS theory.
Furthermore, the fractionally quantized Hall conductivity was computed in
both the gauge theory side and its gravity dual, and it was shown how
the model could be used to compute the topological entanglement 
entropy. \cite{Kitaev:2005dm,Levin:2006zz,Dong:2008ft} 

Our main goal is to show how to realize defects shifting the CS level 
from $(-k)$ to $(-k')$ within this model, and to analyze the
system in detail using holography. The defects are naturally realized
geometrically by $|k-k'|$ D7 branes peeling off the soliton tip to
attach to the AdS boundary along the defect locus. 
Note that the gravity dual can be treated within the supergravity
approximation when $N$ and the 't Hooft coupling
 $\lambda_{\rm 3d}=g_{\rm 3d}^2 N$ 
are large. Therefore, we are dealing with large $N$ strongly coupled regime
of the 3-dimensional gauge theory.\footnote{
In this paper, we treat $k$ to be of $\cO(N^0)$.
This is different from the usual large $N$ analysis
of the CS-theory, in which $k$ is assumed to be of $\cO(N)$.} 

Although the direct relevance of the large $N$ gauge theories to condensed 
matter systems is perhaps questionable, non-Abelian CS theory does have 
known applications to condensed matter theory. 
In the FQH state with filling fraction $\nu=\frac{1}{2}$, for example,
the effective theory of the 
fermionic Moore-Read Pfaffian state can be derived by flux attachment
from the $SU(2)_2$ CS theory 
describing the bosonic Pfaffian state at $\nu=1$
\cite{Fradkin:1997ge}.\footnote{An important 
step to understand these new states was the insight that electrons in a
completely filled Landau level can undergo perturbative p-wave pairing
via the statistical gauge field interaction, and then Bose-Einstein
condense \cite{Greiter:1991vc}. This state is in the same universality
class as the bosonic Pfaffian state \cite{Halperin:1983zz}, and can be
connected to the $\nu=1/2$ state via flux attachment.  
}  
The (non-abelian) edge states of the non-Abelian CS
theory are also known to play an important role in the context of FQHE,
being the edge excitations which carry the 
topologically protected and quantized Hall
response~\cite{Moore:1991ks,Milovanovic:1996nj,Read:1998ed}. In fact,
the derivation of the bulk effective $SU(2)_2$ CS action  
in \cite{Fradkin:1997ge} started from the observation that the edge
theory of the bosonic Pfaffian state at $\nu=1$ is a $SU(2)_2$ Kac-Moody
algebra. 

The contents of this paper can be summarized as follows.
We begin in section~\ref{defects} by introducing 3-dimensional $SU(N)$ 
YM-CS field theory and its level-changing defects, 
followed by its realization by probe branes in section~\ref{braneconfig}.
Section~\ref{Holo} analyses the probe D7($\ol{\rm D7}$) branes
on the gravity side and gives general solutions for the transverse
scalar and the worldvolume gauge fields.  
In section~\ref{correlators}, we perform the holographic renormalization
of the on-shell D7-brane action, and use the renormalized action to
compute holographic correlation functions of defect operators. 
Section~\ref{freeenergy} uses these results to evaluate the free energy
of the D7-brane configuration, revealing a phase transition in the
correlation functions across defects pairs as a function of defect
separation; 
studies in greater detail the question of confinement in
YM-CS from the point of view of the gravity dual; 
and computes the chiral condensate that forms between the chiral fermions
living on adjacent defect/anti-defect pairs.
In section~\ref{finT}, we generalize our considerations to
finite temperature by replacing the AdS soliton with the AdS black hole,
and study the effects of finite temperature on 
the phase transitions of the cross-defect correlators.
Finally in section~\ref{summary}, we summarize our results and
discuss some of the implications and the outlook. In particular,
we point out an interesting relation between the FQHE and 2-dimensional QCD,
and discuss possible applications of our model to FQH physics. Our
notational conventions are summarized in Appendix~\ref{LC}, 
and details of the solutions of the D7-brane equations of motion can be
found in Appendix~\ref{AppB}.

\section{Yang-Mills-Chern-Simons theory and its level-changing defects}
\label{defects}

In this section, we consider 3-dimensional $SU(N)$ 
YM-CS theory defined on a flat spacetime parametrized by $x^\mu$
($\mu=0,1,2$).
We study this theory in the presence of 2-dimensional defects
(domain walls) at which the CS level changes.   
To simplify things, we consider only defects extended along
the $x^\pm\equiv (x^0\pm x^1)/2$ directions%
\footnote{Our convention for the light-cone coordinates is summarized
in Appendix \ref{LC}.}
at fixed values of $y\equiv x^2$ so that the 2-dimensional Lorentz
symmetry is preserved.

\begin{figure}[tp]
\begin{center}
{\includegraphics[height=4cm]{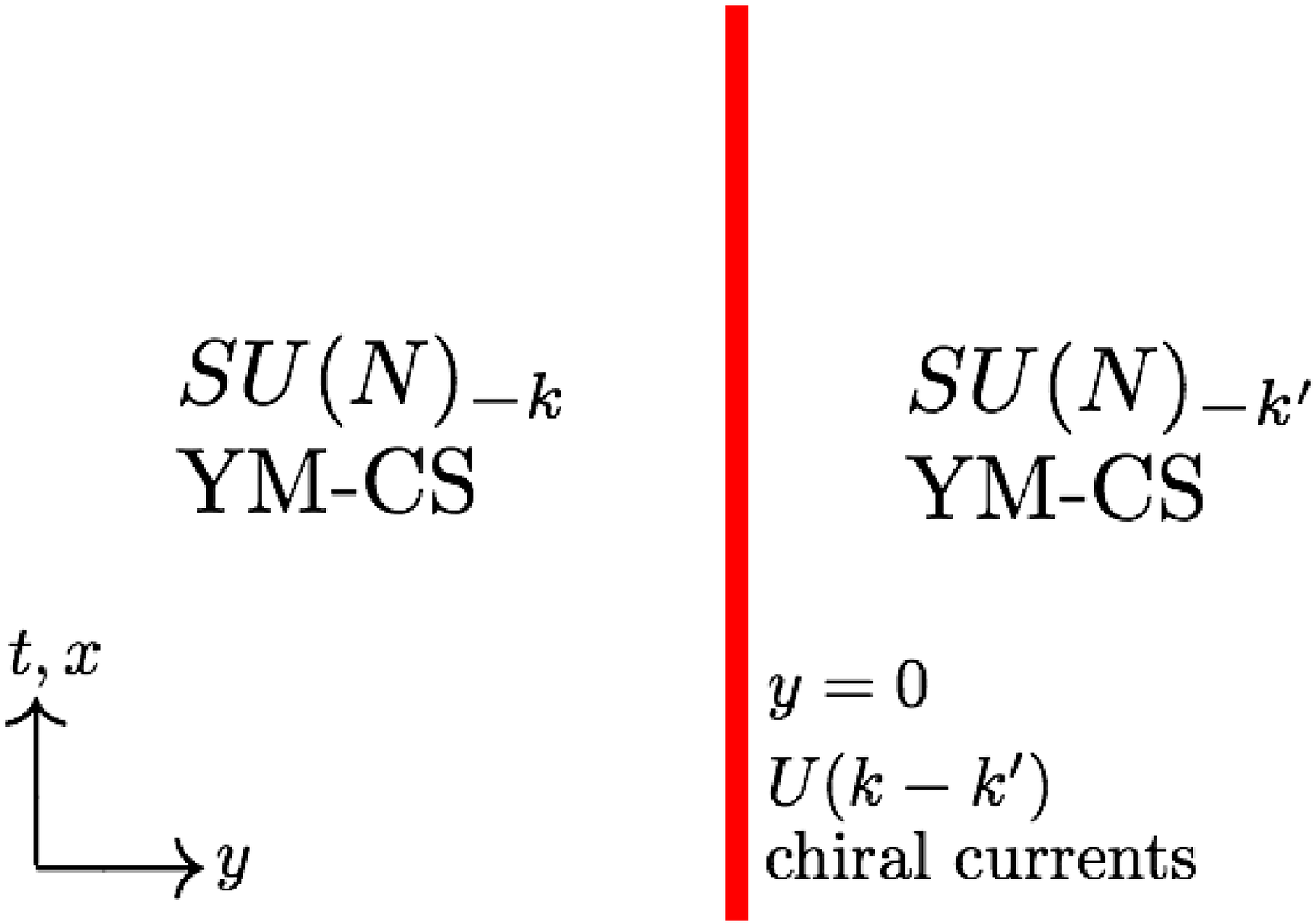}}\hspace{0.5cm}
{\includegraphics[height=4cm]{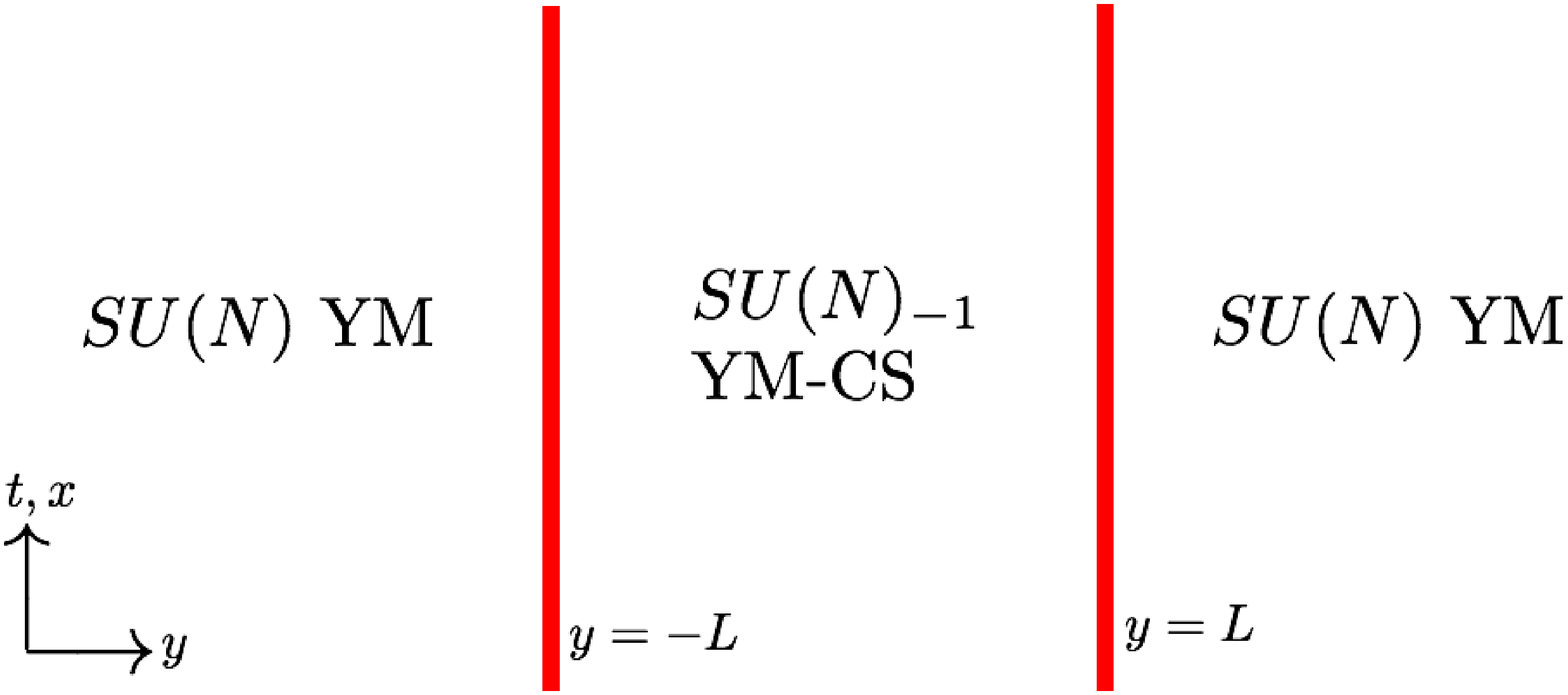}}
\parbox{70ex}{
\caption{Left: A (1+1)-dimensional defect in (2+1)-dimensional $SU(N)$
 YM-CS theory. Chiral fermions on the defect induce
 the jump in the CS level when crossing the defect, and give
 rise to chiral $U(k-k')$ global currents. Right: A more complicated
 configuration in which a YM-CS vacuum with nontrivial
 level can exist between vacua of pure YM theory if the fermions
 on both defects are equal in number but opposite in chirality. 
}
\label{fig:DW1}}
\end{center}
\end{figure}

As a first example, let us consider a defect placed at $y=0$, as
depicted in the left panel of Fig.~\ref{fig:DW1}. 
The level of the CS term is set to be $(-k)$ and
$(-k')$ for the regions $y<0$ and $y>0$, respectively.
We assume that $k$ and $k'$ are integers satisfying $k>k'$.
The Lagrangian for the $SU(N)$ gauge field $A=A_\mu dx^\mu$ is
\begin{eqnarray}
S_{A}=
-\frac{1}{4g_{\rm 3d}^2}\int d^3x\,{\rm Tr}(F^{\mu\nu}F_{\mu\nu})
-\frac{k}{4\pi}\int_{y<0}\omega_3(A)
-\frac{k'}{4\pi}\int_{y>0}\omega_3(A)
\ ,
\label{SA}
\end{eqnarray}
where $\omega_3(A)$ is the Chern-Simons 3-form%
\footnote{We choose the orientation of all $p$-form integrals so that the
integral of $dx^0\wedge dx^1\wedge\cdots$ is positive.}
\begin{eqnarray}
 \omega_3(A)\equiv
\Tr\left(A\wedge dA-\frac{2i}{3}A\wedge A\wedge A\right)\ .
\end{eqnarray}
Note that the CS 3-form transforms as
\begin{eqnarray}
\delta_\alpha \omega_3(A)=d{\rm Tr}(\alpha \, dA)\ ,
\end{eqnarray}
under the infinitesimal gauge transformation
 $\delta_\alpha A=d\alpha-i[A,\alpha]$, and
the action (\ref{SA}) transforms as
\begin{eqnarray}
\delta_\alpha S_A = \frac{k'-k}{4\pi}\int_{y=0}\Tr({\alpha \, dA})\ .
\label{deltaSA}
\end{eqnarray}
Here, we have assumed that the gauge field is continuous at $y=0$, 
and dropped boundary terms at infinity not relevant for our discussion.
In order to have a gauge invariant action,
we put $(k-k')$ negative chirality Weyl fermions $\psi_-^i$
($i=1,2,\cdots, k-k'$),
which transform as the fundamental representation of the gauge
group $SU(N)$, on the 2-dimensional defect at $y=0$. The subscript ``$-$''
of $\psi_-^i$ indicates the chirality of the fermion.
The action for the chiral fermions is
\begin{eqnarray}
 S_\psi=\int_{y=0} d^2x\,\psi_{-i}^\dag (i\del_+ + A_+)\psi_-^i\ ,
\label{Spsi}
\end{eqnarray}
where $\del_\pm\equiv \del_0\pm\del_1$ and $A_\pm\equiv A_0\pm A_1$.
The gauge anomaly induced by the chiral fermion precisely cancels
the anomalous gauge transformation due to the CS term (\ref{deltaSA}),
and the whole system is gauge invariant.\footnote{
Depending on the regularization, local counterterms may also be needed.
} 

In the following sections, we consider operators inserted on the
defects. An important example for the defect operators is the current
operator associated to the $U(k-k')$ global symmetry, which acts on the
chiral fermions on the defect, defined as
\begin{eqnarray}
J^{a+} \equiv \psi_{-j}^\dag (T^a)^j_{~i}\psi_-^i\ ,
\label{J}
\end{eqnarray}
where $T^a$ are the generators of the $U(k-k')$ symmetry.
When the external gauge field $\cA_+$ associated to the $U(k-k')$
symmetry is introduced in the action (\ref{Spsi}), it naturally couples
with the current operator as
\begin{eqnarray}\label{AJ27}
\int_{y=0} d^2x\,\cA^a_+ J^{a+}\ .
\label{AJ}
\end{eqnarray}
As with any gauge field, coupling an external gauge field to
2-dimensional chiral fermions gives rise to a chiral anomaly.
From the point of view of the external $U(k-k')$ symmetry there are $N$
chiral fermions, and as a result
the gauge variation $\delta\cA^a_+ = D_+\Lambda^a$ of the effective
action $S(\cA)$ takes the form
\be
\delta_\Lambda S(\cA) = \frac{N}{4\pi} \int d^2x\,\Lambda^a\del_- \cA_+^a 
\ .
\ee
From $\delta W[\cA] = \int d^2x\, J^{a\mu}(x)\delta \cA^a_\mu(x)$ and  
the relation $J_-=-2J^+$ we obtain the
anomalous current conservation equation\footnote{See the footnote in
p.422 of \cite{Bardeen:1984pm} for a comment on this form of the anomaly
equation.
}
\be
\langle D_+ J_- \rangle = \frac{N}{2\pi}\del_- \cA_+\ .
\label{anomaly}
\ee

Another example is the operator associated with the displacement
of the defect. If we put the defect at $y=\epsilon$ with $|\epsilon|\ll 1$,
the chiral fermions couple with the gauge field evaluated at $y=\epsilon$
and hence the action (\ref{Spsi}) is modified as
\begin{eqnarray}
S_\psi
&\simeq&\int_{y=0} d^2x\left(\psi_{-i}^\dag (i\del_++A_+)\psi_-^i+
\epsilon\,\cO_y\right)
\ ,
\end{eqnarray}
where
\begin{eqnarray}
\cO_y\equiv \psi_{-i}^\dag F_{y+}\psi_-^i\ ,
\label{Oy}
\end{eqnarray}
and $F_{y+}=\del_yA_+ - \del_+A_y - i[A_y,A_+]$.%
\footnote{The easiest way to see this is to work in the $A_y=0$ gauge and 
insert the expansion 
$A_+|_{y=\epsilon}\simeq
A_+|_{y=0}+\epsilon\,\del_yA_+|_{y=0}+\cO(\epsilon^2)$
into (\ref{Spsi}).}
We will also consider a third operator that does not have such a 
straightforward geometric interpretation from the point of view 
of the field theory, the dimension 5 operator
\be
\cO_+ \equiv \psi_{-i}^\dagger F_{+y}F_{+y} \psi^i_- \ .
\label{dim5op}
\ee

When $k<k'$ then, as the coefficient of (\ref{deltaSA}) has the 
opposite sign,
we should introduce $(k'-k)$ chiral fermions with
positive chirality $\psi_+^i$ ($i=1,2,\cdots,k'-k$) on the defect.
Then, the action for the fermions with the source terms is
\begin{eqnarray}
 S_\psi=\int_{y=0} d^2x\,
\left(\psi_{+i}^\dag (i\del_- + A_-)\psi_+^i
+\cA_-^a J^{a-}+\epsilon\cO'_y\right)
\ ,
\label{Spsi2}
\end{eqnarray}
where
\begin{eqnarray}
J^{a-}\equiv\psi_{+j}^\dag(T^a)^j_{~i}\psi_+^i 
\end{eqnarray}
is the current
operator associated with the $U(k'-k)$ symmetry and $\cO'_y$ is defined
as
\begin{eqnarray}
\cO'_y\equiv \psi_{+i}^\dag F_{y-}\psi_+^i\ .
\end{eqnarray}

The generalization to more complicated configurations is
straightforward.
In sections \ref{Holo}--\ref{finT}, we will mainly consider the case with
two defects at $y=-L$ and $y=L$, which change the level of the CS term from
$0$ to $-1$ at $y=-L$ and $-1$ to $0$ at $y=L$ along the $y$ axis. Such
a configuration is depicted in the right panel of Fig.~\ref{fig:DW1}.
The action for the gauge field is
\begin{eqnarray}
S_{A}=
-\frac{1}{4g_{\rm 3d}^2}\int d^3x\,{\rm Tr}(F^{\mu\nu}F_{\mu\nu})
-\frac{1}{4\pi}\int_{-L<y<L}\omega_3(A)\ .
\label{SA2}
\end{eqnarray}
In this case, we put positive and negative chirality fermions at
$y=-L$ and $y=L$, respectively. In the region with $|y|>L$,
the theory is pure YM theory without CS term.

YM theory in 3-dimensions is known to have a mass gap due to
confinement and the Wilson loop exhibits area law behavior.
Pure CS theory on the other hand is a topological theory, 
so that the expectation value of Wilson lines depends only on topology
and is independent of separation.
It is a non-trivial question which behavior arises in the region between
the two defects: the system should be gapped, as the CS term gives the 
gauge field a mass at tree level, but if the gap is sufficiently 
smaller than the confinement scale, the confining behavior may take
over. (See, \emph{e.g.,} \cite{Cornwall:1996jb,Karabali:1999ef} for a
discussion of related issues.)
We address the question of confinement in our system in 
section~\ref{sec:confinement}.

\section{Brane configuration}
\label{braneconfig}

We now turn to the realization of 3-dimensional YM-CS theory with 
level-changing defects by the infrared behavior of 
a brane configuration in string theory.

Consider Type IIB string theory compactified on an $S^1$ of radius
$M_{\rm KK}^{-1}$ and $N$ D3 branes wrapped on it. The D3 brane is
extended along $x^0$, $x^1$, $x^2\equiv y$ and $x^3\equiv\tau$
directions, where $\tau$ parametrizes the $S^1$ direction.
Following \cite{Witten:1998zw}, we impose an
anti-periodic boundary condition on all the fermions along the $S^1$.
This SUSY-breaking boundary condition gives all fermion modes a  
tree level mass of order $M_\mathrm{KK}$.
Quantum corrections then induce masses in the scalar fields, 
lifting them from the infrared spectrum.
The resulting theory is thus expected to flow to 3-dimensional pure
$SU(N)$ YM theory at low energies.\footnote{
Since we do not take into account the singleton degrees of freedom
in our consideration in section \ref{Holo}, the $U(1)$ part of the
$U(N)$ gauge group is dropped. (See, e.g.,
\cite{Witten:1998qj,Aharony:1998qu}
and Appendix B of \cite{Maldacena:2001ss}.)
In any case, the difference between $U(N)$ and $SU(N)$ is not important
in the large $N$ limit.
}
The gauge coupling $g_{\rm 3d}$ for the 3-dimensional YM theory is
identified with
\begin{eqnarray}
g_{\rm 3d}^2=g_s M_{\rm KK}\ ,
\end{eqnarray}
where $g_s$ is the string coupling.

Note that we can safely take the limit $l_s\ra 0$, where
$l_s$ is the string length, so that all the stringy excited states
become infinitely heavy and the couplings to closed strings
vanish. 
To be precise, the resulting theory is not exactly pure 3-dimensional
YM theory, but ${\cal N}=4$ supersymmetric YM theory
compactified on the $S^1$ with SUSY-breaking boundary conditions.
The 3-dimensional pure YM theory is
realized as the massless sector of this configuration, but there are
infinitely many massive Kaluza-Klein (KK) modes associated to the $S^1$.
In principle, in order to make the KK modes
infinitely heavy, we should take the limit
$M_{\rm KK}\ra\infty$ with 
$\lambda_{\rm 3d}\equiv g_{\rm 3d}^2N$ kept finite\footnote{
The typical energy scale in 3-dimensional YM theory is given by
$\lambda_{\rm 3d}$. See \cite{Teper:1998te} for a lattice study of
3-dimensional large N gauge theories.}
by tuning $g_s\ra 0$.
However, in the following sections, we study the holographic description
within the supergravity approximation, which can be trusted only when
$N\gg 1$ and $\lambda_{\rm 3d}\gg M_{\rm KK}$. Therefore, it is not
possible to decouple the Kaluza-Klein modes in the parameter region
we are going to consider.
For this reason, we will keep $M_{\rm KK}$ finite, and mainly consider
the low energy behavior of the theory. We hope that the KK modes
will not alter the qualitative behavior at low energies. 

The CS term is obtained by introducing non-zero RR flux $dC_0$,
where $C_0$ is the RR 0-from field,
along the $S^1$.\cite{Fujita:2009kw}
Recall that the CS term of the D3-brane action has the following
term when $dC_0$ is non-trivial:
\begin{eqnarray}
S_{\rm CS}^{\rm D3}=-
\frac{1}{8\pi^2} \int_{R^3\times S^1} dC_0\wedge\omega_3(A)\ .
\label{D3CS}
\end{eqnarray}
Therefore, when we have
\begin{eqnarray}
 \int_{S^1}dC_0=2\pi k\ ,
\end{eqnarray}
(\ref{D3CS}) gives the CS term at level $(-k)$ and hence we obtain the
brane configuration corresponding to the 3-dimensional $SU(N)$ YM-CS
theory at low energies. 

In order to introduce 2-dimensional defects
with chiral fermions on them, we put D7 branes
extended along $x^0$, $x^1$, $x^4,\cdots, x^9$
directions, as considered in
\cite{Harvey:2007ab,Buchbinder:2007ar,Harvey:2008zz} for the supersymmetric
case. 
When $n$ D7 branes are placed at $y=\tau=0$,
the 3-7 strings (open strings stretched between D3 branes and D7 branes)
give $n$ flavors of chiral fermions as the massless modes.
In addition, since the D7 branes are magnetically charged under RR
0-form field $C_0$, we have the relation
\begin{eqnarray}
 \int_{S^1_-}dC_0- \int_{S^1_+}dC_0
=2\pi n\ ,
\end{eqnarray}
where $S^1_+$ and $S^1_-$ are the $S^1$ in the $\tau$ direction
with $y>0$ and $y<0$, respectively.
Choosing $C_0$ to satisfy 
\begin{eqnarray}
\int_{S^1_+}dC_0 = 2\pi k'\ ,~~
\int_{S^1_-}dC_0 = 2\pi k\ ,
\end{eqnarray}
with $n=k-k'$, the CS term (\ref{D3CS}) becomes
\begin{eqnarray}
S_{\rm CS}^{\rm D3}=
-\frac{k}{4\pi} \int_{y<0} \omega_3(A)
-\frac{k'}{4\pi} \int_{y>0} \omega_3(A)
\ ,
\end{eqnarray}
which agrees with the CS term in (\ref{SA}).
In this way, the first example in section \ref{defects}
is obtained by putting $(k-k')$ D7 branes at $y=\tau=0$.

Similarly, the brane configuration that realizes the theory in (\ref{SA2})
is given by placing a D7 brane and a $\ol{\rm D7}$ brane at $y=L$ and 
$y=-L$, respectively. It is known that the chirality of the massless
fermion in the spectrum of the 3-$\ol 7$ strings is opposite to that of
the 3-7 strings, as required by the anomaly cancellation discussed in
section \ref{defects}.
This configuration is a close analogue to the 
D4-D8-$\ol{\rm D8}$ system used in \cite{Sakai:2004cn} to obtain a
holographic description of QCD. In fact, if we place $N_f$
D7 branes at $(y,\tau)=(0,0)$ and $N_f$ $\ol{\rm D7}$ branes at
 $(y,\tau)=(0,\pi M_{\rm KK}^{-1})$, and T-dualize along the $y$
direction, we obtain the D2-D8-$\ol{\rm D8}$ system considered in
\cite{Rodriguez:2005jr,Yee:2011yn}, which is the 2-dimensional version
of the holographic QCD.

One may question the stability of this
brane configuration. At the least we must make sure that the separation
between the D7 brane and the $\ol{\rm D7}$ brane is larger than the
string length scale $l_s$ so that there is no tachyonic mode in the spectrum
of the open strings connecting the D7 brane and the $\ol{\rm D7}$ brane.
In addition, because the D7 brane and the $\ol{\rm D7}$ brane are
attracted to each other by closed string exchange, 
the asymptotic behavior of the branes may need to be modified to pull
the D7 brane and $\ol{\rm D7}$ branes apart at infinity so as to 
balance the force. 
We will not try to investigate this issue in this paper.
In the following sections, we will only consider the
near horizon limit of the D3-brane background and assume that we can
work in the probe approximation \cite{Karch:2002sh},
in which the backreaction due to the
D7 branes is neglected, with $N$ being much larger than
the number of D7 branes. At least in this limit, it is possible to show
that there are no tachyonic modes in the fluctuations of the D7 brane
in the holographic description.

\section{Holographic description}
\label{Holo}

\subsection{Background geometry}

As mentioned in the previous section, we treat D7 branes as probe branes
embedded in the near horizon geometry corresponding to the D3-brane background.
The background corresponding to the D3 brane considered in section
\ref{braneconfig} is called the AdS soliton background.
The metric as well as the configuration of the other fields
for this background is explicitly known.\cite{Witten:1998zw}%
\footnote{See \cite{Aharony:1999ti} for a review.}
The metric is given by
\begin{eqnarray}
 ds^2=\frac{u^2}{R^2}\left(\eta_{\mu\nu}dx^\mu dx^\nu
+f(u)d\tau^2
\right)+\frac{R^2}{u^2}\frac{du^2}{f(u)}
+R^2 d\Omega_5^2\ ,
\label{metric}
\end{eqnarray}
where $\eta_{\mu\nu}={\rm diag}(-1,1,1)$ $(\mu,\nu=0,1,2)$ is the
3-dimensional Minkowski metric, $d\Omega_5^2$ is the line element of the
unit $S^5$, and
\begin{eqnarray}
 f(u)\equiv 1-\frac{u_0^4}{u^4}\ .
\end{eqnarray}
We also use the coordinates $x^\pm\equiv \half(x^0\pm x^1)$ and
$y\equiv x^2$ as we did in the previous section.
Since $f(u)$ should be positive, the radial coordinate $u$ is restricted
as $u\ge u_0$. The $\tau$ direction is compactified to a circle of radius
$M_{\rm KK}^{-1}$ by the identification
\begin{eqnarray}
 \tau\sim\tau+2\pi M_{\rm KK}^{-1}\ .
\end{eqnarray}
To avoid a conical singularity at $u=u_0$, $M_{\rm KK}$ must be related
to $u_0$ and $R$ by
\begin{eqnarray}
 M_{\rm KK}=\frac{2u_0}{R^2}\ .
\end{eqnarray}
The dilaton field $\phi$ is constant and it is related to the string
coupling $g_s$ as $g_s=e^\phi$. The parameter $R$ in the metric
(\ref{metric}) is related to the string length $l_s=\sqrt{\alpha'}$
and the number of D3 branes $N$ as
\begin{eqnarray}
 R^4=4\pi g_s N l_s^4\ .
\label{R4}
\end{eqnarray}
In addition, the RR 5-form field strength $F_5$ satisfies\footnote{Different
conventions for the normalization of the five-form flux exist in the
literature. Here we follow Appendix A of \cite{Sakai:2004cn}. In another
normalization convention the dimension of   
$\int\limits_{S^5} F_5$ is $[{\rm mass}]^{-4}$, and the flux integral is
quantized in units of $(2\pi \ell_s)^4 g_s$.} 
\begin{eqnarray}
 \int_{S^5} F_5=2\pi N\ .
\label{F5}
\end{eqnarray}

\subsection{Probe D7 brane}
\label{probeD7}

In order to find a consistent D7-brane configuration, we have to solve
the equations of motion for the fields on the D7-brane world-volume.
The D7 branes corresponding to the defects considered in the
previous section are extended along $x^M=(x^0, x^1, u)$
directions and wrapped on the $S^5$.
However, since the coordinate $u$ is bounded from below, it
has to bend in one of the boundary directions.
Let us start with a single D7 brane embedded in the background.
We parametrize the D7-brane world-volume using $x^M$
and the coordinates of the $S^5$.
For simplicity, we only turn on 3-dimensional components $a_M$
($M=0,1,u$) of the 
$U(1)$ gauge field on the D7 brane and consider the configurations that
are uniform along the $S^5$ directions.
The position of the D7 brane in the $(y,\tau)$
space is given by the functions $y=y(x^M)$ and $\tau=\tau(x^M)$,
which are treated as scalar fields on the D7 brane.
The effective action is
\begin{eqnarray}
 S^{\rm D7}=S^{\rm D7}_{\rm DBI}+S^{\rm D7}_{\rm CS}
\end{eqnarray}
with
\begin{eqnarray}
S^{\rm D7}_{\rm DBI}
&=&
-\frac{1}{(2\pi)^7l_s^8g_s}\int dx^0dx^1du\, vol(S^5)R^5
\sqrt{-\det\left(g_{MN}+(2\pi\alpha')f_{MN}
\right)}\ ,
\label{DBI1}
\\
 S^{\rm D7}_{\rm CS}&=&\frac{1}{8\pi^2}\int F_5\wedge a\wedge da\ ,
\label{CS1}
\end{eqnarray}
where $vol(S^5)$ is the volume form of the unit $S^5$,
$g_{MN}$ is the induced metric,
$a=a_Mdx^M$ is the gauge field on the D7 brane
and $f_{MN}\equiv \del_M a_N-\del_N a_M$ is its field strength.
The induced metric can be written explicitly as
\begin{align}
g_{MN} = G_{MN} + G_{ij}\p_M y^i \p_N y^j \ ,
\end{align}
where $y^i=(y,\tau)$ are the embedding functions and
$(G_{MN},G_{ij})$ are part of the background metric read from
(\ref{metric}), whose non-zero components are given as
\begin{eqnarray}
 G_{\mu\nu}=\frac{u^2}{R^2}\eta_{\mu\nu}\ ,
~~G_{uu}=\frac{R^2}{u^2}\frac{1}{f(u)}\ ,
~~G_{yy}=\frac{u^2}{R^2}\ ,~~G_{\tau\tau}=\frac{u^2}{R^2}f(u)\ .
\end{eqnarray}

Integrating over the $S^5$,
the action is reduced to the 3-dimensional DBI-CS action:
\begin{eqnarray}
S^{\rm D7}_{\rm DBI}
&=&
-T_{\rm 3d}\int dx^0dx^1du
\sqrt{-\cG}\ ,
\label{DBI2}
\\
 S^{\rm D7}_{\rm CS}&=&\frac{N}{4\pi}\int a\wedge da
=\frac{N}{8\pi}\int dx^0dx^1du\left(
a_-f_{+u}-a_+f_{-u}-a_uf_{+-}
\right)\ ,
\label{CS2}
\end{eqnarray}
where the effective 3d tension is given by
\begin{align}
T_{\rm 3d}&=\frac{R N}{8\pi(2\pi\alpha')^2}\ ,
\label{T3d} 
\end{align}
and $\cG$ is defined as
\begin{align}
\cG &= \det(\cG_{MN})\ ,
\label{cG} 
\end{align}
where
\begin{align}
\cG_{MN} &= g_{MN}+(2\pi\alpha')f_{MN}\ .
\label{MMN}
\end{align}
Here we have used (\ref{R4}) and (\ref{F5}). Note that the level of the
CS term (\ref{CS2}) is $N$, which we take to be positive. 

The equations of motion for the transverse embedding coordinates
$y^i=(y,\tau)$ and the gauge field $a_M$ take the form
\begin{align}
\p_M \left(\sqrt{-\cG}G_{ij}\cG^{MN}_S\p_N y^j\right) &= 0 \ ,
\label{yEOM}
\\
(2\pi\alpha')T_{\rm 3d}\p_M\left(\sqrt{-\cG}\cG_A^{MN}\right) 
&= \frac{N}{4\pi}\epsilon^{NPQ}f_{PQ}\ .
\label{aEOM}
\end{align}
Here $\cG_S^{MN}$ and $\cG_A^{MN}$ are defined as
\begin{align}
\cG_{\rm S}^{MN} & = \half(\cG^{MN}+\cG^{NM})\ ,
\label{MS}
\\
\cG_{\rm A}^{MN}& = \half( \cG^{MN}- \cG^{NM})\ ,
\label{MA}
\end{align}
which are respectively the symmetric and antisymmetric parts  
of the inverse matrix $\cG^{MN}$ of $\cG_{MN}$, \emph{i.e.} 
$\cG^{MN}\cG_{NP} = \delta^M_P$. 
(See Appendix \ref{eomDBI-CS} for more details.)
In Appendix \ref{Sol0}, we summarize the solutions of these equations 
in the case that $y^i$ and $f_{MN}$ depend only on $u$.
If we further assume that the gauge field $a_M$ depends only on $u$,
the most general solution is
\footnote{With this assumption, $a_u$ does not appear in the equations
of motion and can be an arbitrary function of $u$.}
\begin{eqnarray}
y(u)&=&y_0+ c_y\int^u_{u_{\rm min}} du'\frac{R^5}{F(u')}\ ,
\label{y}
\\
\tau(u)&=&\tau_0+c_\tau\int_{u_{\rm min}}^u du'
\frac{R^5}{f(u')F(u')}\ ,
\\
a_\pm(u)
&=&a_\pm^{(0)}
\pm\frac{c_\pm}{8\pi\alpha'}
\exp\left(
\mp 4\int_{u_{\rm min}}^u du'\frac{u'^4}{F(u')}
\right)\ ,
\label{apm}
\end{eqnarray}
where $y_0$, $c_y$, $\tau_0$, $c_\tau$, $a_\pm^{(0)}$, $c_\pm$
and $u_{\rm min}$ are constants, and
\begin{eqnarray}
F(u)\equiv \sqrt{
u^4 f(u)\left(u^6+u^4c_+ c_--R^6c_y^2-\frac{R^6c_\tau^2}{f(u)}
\right)}\ .
\label{Fu}
\end{eqnarray}

To see what the solution looks like, consider the case with
$\tau_0=c_\tau=a_\pm^{(0)}=c_\pm=0$ and $c_y>0$. Then, (\ref{y}) becomes
\begin{eqnarray}
 y(u)=y_0+R^2 u_*^3\int_{u_{\rm min}}^u
  \frac{du'}{\sqrt{(u'^4-u_0^4)(u'^6-u_*^6)}} \ ,
\label{eq:y}
\end{eqnarray}
where $u_* \equiv R c_y^{1/3} $. 

Na\"ively, for a given choice of asymptotic boundary conditions on $y(u)$
as $u\to\infty$ there are two distinct branches of solutions, one with
$u_* \ge u_0$ and the other with $u_* < u_0$.
However, the second branch corresponds to a D7 brane whose two ends
asymptotically approach opposite sides of the $\tau$ circle.
To better understand this solution, define coordinates $(\rho,\theta)$
on the $(u,\tau)$ plane by the identifications
\be
\rho^2 = \frac{R^2u^2}{4u_0^2}f(u)\ ,
\qquad
\theta = \frac{2u_0}{R^2}\tau\ .
\ee
Note that $\theta$ has periodicity $2\pi$.
The 5-dimensional metric takes on the form
\be
ds^2_5 = \frac{u^2(\rho)}{R^2}\eta_{\mu\nu}dx^\mu dx^\nu +
\frac{4u_0^2/u^2(\rho)}{\left(1+u_0^4/u^4(\rho)\right)^2}\,d\rho^2 
+ \rho^2\, d\theta^2\ .
\ee
We further introduce coordinates $(v,w)$ by
\be
v = \rho\, \cos\theta\ , \qquad w = \rho\, \sin\theta \ ,
\ee
in which the asymptotic region corresponds to $\rho^2=v^2+w^2\to\infty$.
The solutions we consider have $\tau=0$, which in the new coordinates
is $w=0$.
We thus wish to solve for $y$ as a function of $v$.
The differential relation now becomes 
\be
\frac{dy}{dv} = \frac{2Ru_*^3u_0}{u^2(v)(1+u_0^4/u^4(v))\sqrt{u^6(v)-u_*^6}}\ .
\ee
In the case $u_*<u_0$, the resulting brane profile $v(y)$ does not have
a turning point. 
Instead, the solutions behave as $v(y)\to \pm\infty$ as $y\to\pm\infty$.
Referring to our original coordinate system, we see that $w=0$, $v<0$
corresponds to the angular position $\theta=\pi$.
Thus the branch $u_*<u_0$ corresponds to a defect with $\theta\to 0$ as
$v\to\infty$, and $\theta\to\pi$ as $v\to-\infty$. 

\begin{figure}[ht]
\begin{center}
\begin{picture}(200,120)(0,0)
\put(5,0){\includegraphics[scale=0.7]{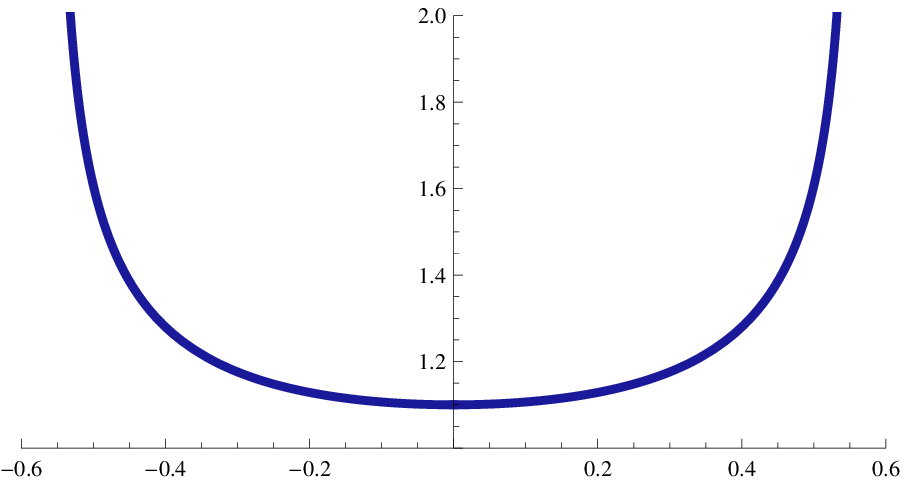}}
\put(195,8){\makebox(0,0){$y$}}
\put(96,103){\makebox(0,0){$u$}}
\put(113,25){\vector(-2,-1){14}}
\put(120,25){\makebox(0,0){$u_*$}}
\end{picture}
\parbox{80ex}{
\caption{A plot of the solution (\ref{sol1}) with $u_0=R=1$ and
 $u_*=1.1$.
}
\label{Usol}}
\end{center}
\end{figure}

We will therefore focus on the case $u_*\ge u_0$. 
It is convenient to choose
$u_{\rm min}=u_*$ and $y_0=0$. This solution makes sense for $u\ge u_*$
and terminates at $u=u_*$. Actually, $u=u_*$ is a turning point and the
solution is smoothly connected to the solution obtained by flipping the
sign of $c_y$ as
\begin{eqnarray}
 y(u)=\pm R^2 u_*^3\int_{u_*}^u
  \frac{du'}{\sqrt{(u'^4-u_0^4)(u'^6-u_*^6)}} \ .
\label{sol1}
\end{eqnarray}
The solution is U-shaped as depicted in Fig.~\ref{Usol}.
The asymptotic value of $y$ is given by $y({u=\infty})=\pm L$ with
\begin{eqnarray}
L\equiv \lim_{u\ra\infty}|y(u)|=
 R^2u_*^3\int_{u_*}^\infty\frac{du}{\sqrt{(u^4-u_0^4)(u^6-u_*^6)}}
\  .
\label{L}
\end{eqnarray}

It is often convenient to use a coordinate that can smoothly parametrize
the D7-brane world-volume around $u=u_*$. One way to do this is to
introduce a coordinate $z$ related to $u$ by
\begin{eqnarray}
 u^2=u_*^2(1+z^2)\ .
\label{z}
\end{eqnarray}
Then, (\ref{sol1}) can be written as
\begin{eqnarray}
 y(z)= R^2 u_*\int_{0}^z
  \frac{dz}{\sqrt{(u_*^4(1+z^2)^2-u_0^4)(1+z^2)(3+3z^2+z^4)}} \ ,
\label{sol2}
\end{eqnarray}
which is valid for $-\infty<z<\infty$.

The configuration given by the solution (\ref{sol1}) corresponds to the
case with a D7 brane and $\ol{\rm D7}$ brane placed at $y=L$ and $y=-L$,
respectively, considered in section \ref{braneconfig}.
As explained around (\ref{D3CS}), the CS level 
for the $SU(N)$ YM-CS theory is given by the integration of the RR 1-form
field strength $dC_0$ along the $S^1$ parametrized by $\tau$.
In the holographic description, it corresponds to minus the number of D7 branes
penetrating the $(u,\tau)$ plane. Therefore, the configuration
given by (\ref{sol1}) (or (\ref{sol2}))
corresponds to the $SU(N)$ YM-CS theory with the level $(-1)$
CS term in the region $-L<y<L$ considered around (\ref{SA2}).

The $L$ defined in (\ref{L}) is a monotonically decreasing function of $u_*$
and it diverges in the limit $u_*\ra u_0$. In this limit, 
the two defects are pushed to infinity and the D7 brane is
placed at $u=u_0$.
This is the configuration corresponding to $SU(N)$
YM-CS theory without a defect considered in \cite{Fujita:2009kw}.
When $k$ D7 branes are placed at
$u=u_0$, it corresponds to the $SU(N)$ YM-CS theory at level $(-k)$.
Interestingly, as pointed out in \cite{Fujita:2009kw}, 
the world-volume theory realized on the D7 branes is a $U(k)$ DBI-CS
theory at level $N$, which implies the level-rank duality
of CS theory at low energy. 
Our construction should therefore give us insight into how level-rank 
duality acts on level-changing defects.

\begin{figure}[ht]
\begin{center}
\begin{picture}(250,135)(0,0)
\put(5,0){\includegraphics[scale=0.9]{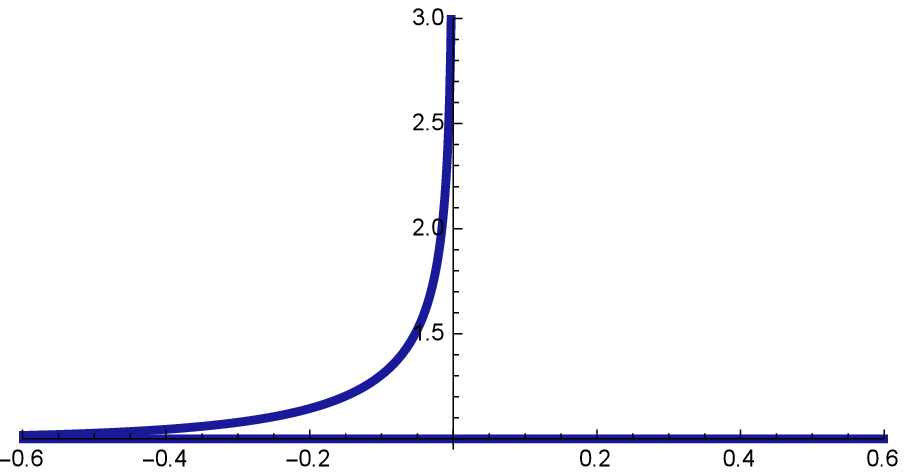}}
\put(245,10){\makebox(0,0){$y$}}
\put(123,126){\makebox(0,0){$u$}}
\put(143,95){\vector(-1,0){16}}
\put(195,95){\makebox(0,0){$(k-k')$ D7 branes}}
\put(40,27){\makebox(0,0){$k$ D7 branes}}
\put(180,22){\makebox(0,0){$k'$ D7 branes}}
\end{picture}
\parbox{70ex}{
\caption{A D7-brane configuration with a single defect.
($u_0=R=1$)
}
\label{single}}
\end{center}
\end{figure}

A configuration with a single defect is obtained by pushing one of the
two defects in the U-shaped solution (\ref{sol1}) to infinity.
It can be achieved by taking a limit
$u_*\ra u_0$, while keeping one defect at a finite position by adjusting
$y_0$ appropriately. For example, a solution corresponding to a defect
placed at $y=0$ is given by
\begin{eqnarray}
 y(u)=- R^2 u_0^3\int_u^\infty
  \frac{du'}{\sqrt{(u'^4-u_0^4)(u'^6-u_0^6)}} \ .
\end{eqnarray}
If there are $(k-k')$ D7 branes satisfying this equation and,
in addition, $k'$ D7 branes placed at $u=u_0$,
we have $k$ and $k'$ D7 branes in $y<0$ and $y>0$, 
respectively, as depicted in Fig. \ref{single}.
(Here we have assumed $0<k'<k$.)\footnote{
In our conventions, a single D7 brane at the tip of the AdS soliton
induces a CS level $(-1)$. Hence positive CS levels
require negative numbers of D7 branes, \emph{i.e.} 
 $\overline{\rm D7}$ branes.}  
This configuration corresponds to the setup described around
(\ref{SA}). Note that the gauge group on the D7-brane world-volume
is $U(k)$ at $y\ra -\infty$, where $k$ D7 branes are placed
at the tip of the AdS soliton ($u=u_0$). This gauge group is
Higgsed to $U(k')\times U(k-k')$ by peeling 
$(k-k')$ D7 branes off from the tip in $-\infty<y<0$.
The $U(k-k')$ factor becomes the global symmetry
on the defect at $y=0$, where the $(k-k')$ D7 branes
reach the boundary $u\ra\infty$. The $U(k')$ factor, on the other hand,
remain intact and continues to be the gauge group on the D7 brane
world-volume in $y>0$. In this way, the level-changing defect
at $y=0$ is mapped to the rank-changing defect
on the D7-brane world-volume, as suggested by the level-rank duality.

Let us next examine solutions with non-trivial gauge fields
on the D7 brane.
Here, we consider the U-shaped solution with $c_\pm\ne 0$ and
 $c_\tau=0$.
In this case, the turning point $u_*$ is related to $c_y$ by
\begin{eqnarray}
c_y^2=\frac{u_*^6}{R^6}\left(1+\frac{c_+ c_-}{u_*^2} \right)\ .
\end{eqnarray}
If we use the coordinate $z$ introduced in (\ref{z}), the solution
(\ref{y})--(\ref{apm}) becomes
\begin{eqnarray}
y(z)&=&
\frac{R^2}{u_*}\sqrt{1+\frac{c_+c_-}{u_*^2}}\int_0^z
\frac{dz'}{\wt F(z')}
\ ,
\label{y2}
\\
a_\pm(z)&=&a_\pm^{(0)}
\pm\frac{c_\pm}{8\pi\alpha'}
f^{(\mp)}(z)
\ ,
\label{apm2}
\end{eqnarray}
where we have set $y_0=0$, and defined
\begin{eqnarray}
\wt F(z)\equiv
\sqrt{(1+z^2)\left((1+z^2)^2-\frac{u_0^4}{u_*^4}\right)
\left(3+3z^2+z^4+(2+z^2)\frac{c_+c_-}{u_*^2}\right)}\ ,
\end{eqnarray}
and
\begin{eqnarray}
f^{(\pm)}(z)
&\equiv& \exp\left(\pm 4\int_{0}^z dz'
\frac{(1+z'^2)^2}{\wt F(z')}\right)\ .
\label{fpm}
\end{eqnarray}
This function (\ref{fpm}) satisfies
\begin{eqnarray}
 f^{(\pm)}(z)=f^{(\mp)}(-z)=\frac{1}{f^{(\mp)}(z)}\ ,
\end{eqnarray}
and the asymptotic behavior is
\begin{eqnarray}
f^{(+)}(z)&\simeq&
\frac{e^{\xi}}{u_*^4}\left(
u^4+c_+c_- u^2+\frac{c_+^2c_-^2}{8}
-\frac{u_0^4}{2}
\right)+\cO(u^{-2})\ ,
\\
f^{(-)}(z)&\simeq&
e^{-\xi}\frac{u_*^4}{u^4}+\cO(u^{-6})\ ,
\end{eqnarray}
for $z=\sqrt{u^2/u_*^2-1}\ra +\infty$,
where we have defined
\begin{eqnarray}
\xi \equiv
4\int_{u_*}^\infty du\left(
\frac{u^4}{F(u)}-\frac{1}{u}
\right)
= 4\int_0^\infty dz\left(\frac{(1+z^2)^2}{\wt F(z)}-\frac{z}{1+z^2}\right)
 \ .
\label{xi}
\end{eqnarray}

The solution (\ref{y2})--(\ref{apm2}) behaves as
\begin{eqnarray}
y(u)&\simeq&
 L-\frac{R^2}{4}\sqrt{1+\frac{c_+c_-}{u_*^2}}\frac{u_*^3}{u^4}+\cO(u^{-6})
\ ,
\label{asymy}
\\
 a_+(u)&\simeq& a_+^{(0)}+\frac{c_+ e^{-\xi}}{8\pi\alpha'}
\frac{u_*^4}{u^4} +\cO(u^{-6})\ ,
\label{asymap}
\\
 a_-(u)
&\simeq& \frac{c_- e^{\xi}}{8\pi\alpha'u_*^4}
\left(u^4+c_+c_-u^2+
\frac{c_+^2c_-^2}{8}-\frac{u_0^4}{2}\right)
+a_-^{(0)}
 +\cO(u^{-2})\ ,
\label{asymam}
\end{eqnarray}
for $z\ra +\infty$, and
\begin{eqnarray}
y(u)&\simeq&
-L+\frac{R^2}{4}\sqrt{1+\frac{c_+c_-}{u_*^2}}\frac{u_*^3}{u^4}+\cO(u^{-6})\ ,
\\
 a_+(u)
&\simeq& \frac{c_+ e^{\xi}}{8\pi\alpha'u_*^4}
\left(u^4+c_+c_-u^2+\frac{c_+^2c_-^2}{8}-\frac{u_0^4}{2}\right)
+a_+^{(0)}
 +\cO(u^{-2})\ ,
\label{asymap2}
\\
a_-(u)&\simeq& a_-^{(0)}+\frac{c_- e^{-\xi}}{8\pi\alpha'}
\frac{u_*^4}{u^4} +\cO(u^{-6})\ ,
\label{asymam2}
\end{eqnarray}
for $z\ra -\infty$,
where
\begin{eqnarray}
L=
\frac{R^2}{u_*}\sqrt{1+\frac{c_+c_-}{u_*^2}}\int_0^\infty
\frac{dz}{\wt F(z)}\ .
\label{L2}
\end{eqnarray}

\section{Operators on the defects}
\label{correlators}

\subsection{Defect mode/operator map}
\label{map}

The map between operators on the defect and fields on the D7 brane when 
supersymmetry is unbroken was analyzed in \cite{Harvey:2008zz}.
Since the asymptotic behavior of the background metric (\ref{metric})
and the D7-brane configuration is the same as that used in
\cite{Harvey:2008zz}, the results can be applied to our system.

Note that the brane configuration considered in section
\ref{braneconfig} is invariant under the $SO(6)$ symmetry that rotates
the $S^5$.
Since the 3-dimensional gauge field and the fermions on the 
defects are all singlets of $SO(6)$, we are interested in the operators 
that are invariant under the $SO(6)$ symmetry.
There are four $SO(6)$ invariant defect operators, denoted
here as $J_-$, $\cO_y$,  $\cO_\tau$ and $\cO_+$,\footnote{
They correspond to $\cO^{(0)}$,
$Q_iQ_j\cO^{(1)}$, $Q^{\dag i}Q^{\dag j}\cO^{(1)}$, and
$Q_iQ_jQ^{\dag k}Q^{\dag l}\cO^{(2)}$
in Table 4 of \cite{Harvey:2008zz}.
Note that our conventions for the light-cone coordinates are the
reversed relative to this reference, $x^\pm_{\rm here} = x^{\mp}_{\rm
there}$.}
corresponding to $SO(6)$ invariant modes on the D7 brane at $y\ra +L$ in
the brane configuration in section \ref{probeD7}.
As the notation suggests, $J_-$ is (the $U(1)$ part of) the current
operator considered in (\ref{J}). 
Keeping only the gauge field $A_\mu$ ($\mu=0,1,2$) and the defect
fermion $\psi_-$ in the analysis of \cite{Harvey:2008zz}, it can be
shown that $\cO_y$ corresponds to the operator defined in (\ref{Oy}),
and $\cO_+$ corresponds to the dimension~5 operator (\ref{dim5op}). 
Since
$\cO_\tau$ involves $A_\tau$ (the $\tau$ component of the gauge field on
the D3 brane) or the derivative with respect to $\tau$, we will not
consider it in the following.
The conformal dimension of these operators and the corresponding fields
on the D7 brane are listed in Table \ref{operator}.
\begin{table}[h]
\centering
\begin{tabular}[t]{|c|c|c|c|}
\hline
operator&$\Delta$&source&vev\\
\hline
$\cO_y$ &3& $y$&$y$ \\
$J_-$ &1& $a_+$&$a_-$ \\
$\cO_+$ &5 & $a_-$&$a_+$ \\
\hline
\end{tabular}
\caption{Defect mode/operator map \label{operator}}
\end{table}

As suggested in this table, the sources of the operators $\cO_y$, $J_-$
and $\cO_+$ correspond to the leading components of the asymptotic
expansion of the fields $y$, $a_+$ and $a_-$, respectively.
$\Delta$ in the table is the conformal dimension of these operators.
Note that the leading terms for $y$, $a_+$ and $a_-$
in (\ref{asymy})--(\ref{asymam}) are $\cO(u^0)$, $\cO(u^0)$
and $\cO(u^4)$, respectively.
Because the dimensions of the bulk objects $u$, $y$ and $a_\pm$ 
under rescalings of the boundary
are $+1$, $-1$ and $+1$, respectively, these asymptotic behaviors are 
consistent with the conformal dimensions of the sources for $\cO_y$, 
$J_-$ and $\cO_+$, which are $-1$, $1$ and $-3$, respectively. 
The correlation functions of these operators can be computed by the
variation of the on-shell action with respective to the sources.
The vacuum expectation values $\VEV{\cO_y}$, $\VEV{J_-}$ and
$\VEV{\cO_+}$ are contained in the
$\cO(u^{-4})$, $\cO(u^{0})$ and $\cO(u^{-4})$ terms
in the asymptotic expansions of $y$, $a_-$, and $a_+$,
respectively. We will examine them in the following subsections.

The argument above can also be applied to the other defect at $y\ra -L$,
and the table corresponding to Table \ref{operator} is obtained by the
replacement: $\cO_y \ra \cO_y'$, $J_-\ra J_+$, $\cO_+\ra \cO_-$
and $a_\pm \ra a_\mp$.

\subsection{Holographic renormalization}

Following the general prescription of AdS/CFT correspondence
\cite{Witten:1998qj,Gubser:1998bc}, the correlation functions are
obtained by varying the
on-shell action. We are mostly interested in the correlation functions
of the defect operators. For this purpose, one should evaluate the
on-shell DBI-CS action including the counterterms that cancel the
divergences in the on-shell action to make it well-defined.\footnote{
See, e.g. \cite{Skenderis:2002wp}
for a review of the holographic renormalization.}
The general procedure of the holographic renormalization for our system
turns out to be very complicated.\cite{futurework}
In order to avoid such complications,
we restrict our consideration to the $x^\pm$ independent
configurations, which still contain interesting information as we
will soon show.

\subsubsection{Counterterms and the on-shell action}

First, let us find the counterterms needed to cancel the
divergences.
Here, we consider the solution (\ref{y2})--(\ref{apm2}).
In this section, we allow the gauge field to depend on $x^\pm$,
while assuming that its field strength is  $x^\pm$ independent.
Then, the constant part $a_\pm^{(0)}$ in (\ref{apm2}) and
$a_u^{(0)}\equiv a_u$ can be promoted to a flat connection satisfying
\begin{eqnarray}
\del_M a_N^{(0)}-\del_N a_M^{(0)}=0\ .
\label{flatcon}
\end{eqnarray}
As we will see in section \ref{bdrycond},
$a_\pm^{(0)}$ should not diverge at $z\ra\pm\infty$
so that the boundary condition is consistent
with the variational principle.

Inserting the solution (\ref{y2})--(\ref{apm2})
into the DBI-CS action (\ref{DBI2})--(\ref{CS2}), 
and using (\ref{detM2}), the on-shell action is
evaluated as
\begin{eqnarray}
S_{\rm DBI}^{\rm o.s.}&=&- T_{\rm 3d}\int d^2 x
\int_{z_-}^{z_+}dz\,\frac{u_*^2(1+z^2)^3}{R\wt F(z)}\ ,
\label{onshellDBI}
\\
S_{\rm CS}^{\rm o.s.}&=&
\frac{N}{8\pi}\int d^2x \int_{z_-}^{z_+} dz\,
\frac{(1+z^2)^2}{\wt F(z)}
\left(
\frac{a_-^{(0)}c_+}{2\pi\alpha'} f^{(-)}(z)
-
\frac{a_+^{(0)}c_-}{2\pi\alpha'} f^{(+)}(z)
-\frac{c_+c_-}{2(2\pi\alpha')^2}
\right)\ ,
\label{onshellCS}
\end{eqnarray} 
where $z_\pm$ is the UV cut-off introduced
to regularize the divergence at $z\ra\pm\infty$.
It can be easily seen that both (\ref{onshellDBI}) and (\ref{onshellCS})
are divergent both in the limit $z_+\ra\infty$ and $z_-\ra -\infty$.
The divergent terms at $z_+\ra\infty$ are
\begin{eqnarray}
S_{\rm DBI}^{\rm o.s.}|_{z_+\ra\infty}&\simeq& -T_{\rm 3d}\int d^2 x
\frac{1}{2R}
\left(u_{\rm max}^2-c_+c_-
\log u_{\rm max}\right) +\mbox{finite}\ ,
\label{SDBIdiv1}
\\
 S_{\rm CS}^{\rm o.s.}|_{z_+\ra\infty} &\simeq&
\frac{N}{8\pi}\int d^2 x
\left(
-a_+^{(0)}\frac{c_- e^\xi}{8\pi\alpha'u_*^4}
\left(u_{\rm max}^4+c_+c_-u_{\rm max}^2
\right)
-\frac{c_+c_-}{2(2\pi\alpha')^2}\log u_{\rm max}
\right) +\mbox{finite}\ ,
\nn\\
\label{SCSdiv1}
\end{eqnarray}
where $u_{\rm max}\equiv u_*\sqrt{1+z_+^2}$. 
Similarly, the divergent terms at
$z_-\ra -\infty$ are
\begin{eqnarray}
S_{\rm DBI}^{\rm o.s.}|_{z_-\ra-\infty}&\simeq& -T_{\rm 3d}\int d^2 x
\frac{1}{2R}
\left(u_{\rm max}^2-c_+c_-
\log u_{\rm max}\right) +\mbox{finite}\ ,
\label{SDBIdiv2}
\\
 S_{\rm CS}^{\rm o.s.}|_{z_-\ra-\infty} &\simeq&
\frac{N}{8\pi}\int d^2 x
\left(
-a_-^{(0)}\frac{c_+ e^\xi}{8\pi\alpha'u_*^4}
\left(u_{\rm max}^4+c_+c_-u_{\rm max}^2
\right)
-\frac{c_+c_-}{2(2\pi\alpha')^2}\log u_{\rm max}
\right) +\mbox{finite}\ ,
\nn\\
\label{SCSdiv2}
\end{eqnarray}
with $u_{\rm max}\equiv u_*\sqrt{1+z_-^2}$. 

The relation (\ref{T3d}) implies that the log divergent terms in
$S_{\rm DBI}$ and $S_{\rm CS}$ cancel each other. In order to
cancel the $\cO(u_{\rm max}^2)$ term in the DBI action, we add
a counterterm of the form
\begin{eqnarray}
S_{\gamma\pm}\equiv \frac{T_{\rm 3d}R}{2}\int_{z=z_\pm}
\hspace{-2ex} d^2 x\sqrt{-\gamma}\ ,
\label{Sgamma}
\end{eqnarray}
where $\gamma=\det(\gamma_{ab})$ ($a,b=0,1$) is the determinant of the
induced metric on the 2-dimensional boundary defined at $z=z_\pm$. In
fact, the induced metric is given as
\begin{eqnarray}
 \gamma_{ab}=g_{ab}|_{z=z_\pm}=\frac{u_{\rm max}^2}{R^2}\eta_{ab}\ ,
\end{eqnarray}
and the counterterm (\ref{Sgamma}) precisely cancel
the $\cO(u_{\rm max}^2)$ terms in (\ref{SDBIdiv1}) and (\ref{SDBIdiv2}).
The $\cO(u_{\rm max}^4)$ and $\cO(u_{\rm max}^2)$ terms in the CS term
are canceled by a counterterm of the form
\cite{Elitzur:1989nr,Davis:2007ka,Jensen:2010em,Yee:2011yn}
\begin{eqnarray}
S_{a\pm}&\equiv&\frac{N}{8\pi} \int_{z=z_\pm}
\hspace{-2ex} d^2x\sqrt{-\gamma}\,\gamma^{ab} a_aa_b
=-\frac{N}{8\pi} \int_{z=z_\pm}\hspace{-2ex}d^2x\,a_+a_-\ .
\label{Saa}
\end{eqnarray}
The on-shell value of this counterterm is
\begin{eqnarray}
&&S^{\rm o.s.}_{a\pm}\nn\\
&\simeq&
-\frac{N}{8\pi} \int_{z=z_\pm}d^2x
\left(
\mp a_\pm^{(0)}\frac{c_\mp}{8\pi\alpha'}
f^{\pm}(u_{\rm max})
+a_+^{(0)}a_-^{(0)}
-\frac{c_+c_-}{(8\pi\alpha')^2R^2}
\right)
\label{onshellSa}
\\
&\simeq&
\frac{N}{8\pi} \int_{z=z_\pm}\hspace{-2ex}d^2x
\Bigg(
a_\pm^{(0)}\frac{c_\mp}{8\pi\alpha'}
\frac{e^{\xi}}{u_*^4}\left(
u_{\rm max}^4+c_+c_- u_{\rm max}^2
+\frac{c_+^2c_-^2}{8}
-\frac{u_0^4}{2}
\right)
-a_+^{(0)}a_-^{(0)}
+\frac{c_+c_-}{(8\pi\alpha')^2}
\Bigg)\ .
\nn\\
\end{eqnarray}
Here, we omitted the terms that vanish at
$z_\pm\ra\pm\infty$ ($u_{\rm max}\ra\infty$). This counterterm
cancels the divergent terms in
(\ref{SCSdiv1}) and (\ref{SCSdiv2}).
Therefore, the total action we consider is\footnote{There are more
counterterms needed to cancel the divergence for the general solution
that has non-trivial $x^\pm$ dependence.\cite{futurework}
}
\begin{eqnarray}
 S_{\rm total}\equiv S_{\rm DBI}+S_{\rm CS}+S_{\gamma+}+S_{\gamma-}
+S_{a+}+S_{a-}\ .
\end{eqnarray}
Collecting the expressions (\ref{onshellDBI}),
 (\ref{onshellCS}), (\ref{Sgamma}) and (\ref{onshellSa}),
the on-shell action is evaluated as
\begin{eqnarray}
 S^{\rm o.s.}_{\rm total}
&=&-
\frac{Nu_*^2}{8\pi(2\pi\alpha')^2}\int d^2 x
\int_{-\infty}^{+\infty}dz
\left(\frac{(1+z^2)^2}{\wt F(z)}\left(1+z^2+\frac{c_+c_-}{2 u_*^2}\right)
-|z|\right)
\nn\\
&&
-\frac{N}{8\pi}\int d^2x
\left((a_+^{(0)}a_-^{(0)})\big|_{z\ra+\infty}
+(a_+^{(0)}a_-^{(0)})\big|_{z\ra-\infty}
-\frac{2c_+c_-}{(8\pi\alpha')^2}
-\frac{u_*^2}{(2\pi\alpha')^2}
\right)\ .
\label{ostotal}
\end{eqnarray}
Here, we have used the relations
\begin{eqnarray}
&& \del_z f^{(\pm)}(z)=\pm 4\frac{(1+z^2)^2}{\wt F(z)}f^{(\pm)}(z)\ ,\\
&& u_{\rm max}^2=u_*^2\left(\int_{z_-}^{z_+}dz\,|z|+1\right)\ ,
\end{eqnarray}
for $z_+=-z_-=\sqrt{u_{\rm max}^2/u_*^2-1}$,
and dropped the terms proportional to
\begin{eqnarray}
\int d^2x \int_{z_-}^{z_+}dz\,\del_z a_\pm^{(0)}c_\mp f^{(\pm)}(z)\ ,
\end{eqnarray}
because these terms are total derivative in the $x^\pm$ direction,
using the flatness condition (\ref{flatcon}).

\subsubsection{Boundary conditions}
\label{bdrycond}

Motivated by the asymptotic behavior of the solutions 
(\ref{asymy})--(\ref{asymam2}) 
and the consideration in section \ref{map},
we impose the boundary condition for the gauge field as 
\begin{eqnarray}
 a_\pm(x^a,z)\ra \cA_\pm(x^a)\ ,~~
 a_\mp(x^a,z)\frac{R^8}{u^4}
\ra \cC_\mp(x^a)\ ,~~~(z\ra \pm\infty)\ ,
\label{abdry}
\end{eqnarray}where $a=0,1$, and $\cA_\pm$ and $\cC_\mp$ are fixed values.
These $\cA_\pm$ and $\cC_\mp$ are interpreted as the sources
that couple to the operators $J_{\mp}$ and $\cO_{\pm}$ on the defects
placed at $y=y(z)|_{z\ra\pm\infty}$, respectively.
Similarly, the boundary condition for the scalar field $y$ is
\begin{eqnarray}
 y(x^a,z)\ra \cY^{(\pm)}(x^a)\ ,~~~(z\ra\pm\infty)\ .
\label{ybdry}
\end{eqnarray}
$\cY^{(\pm)}(x^a)$ is the source of $\cO_y$ and $\cO_y'$.

Let us check that our solution (\ref{y2})--(\ref{apm2}) and
the boundary conditions (\ref{abdry}) and (\ref{ybdry}) are
consistent with the variational principle including the contributions
from the boundaries. The variation of the action
gives surface terms as
\begin{eqnarray}
\delta S_{\rm DBI}&=&\mbox{(EOM)}
-T_{\rm 3d}
\int d^2 x\,\left[
\sqrt{-{\cG}}\left(
\frac{u^2}{R^2}\cG_{\rm S}^{u N}\del_{N} y\,\delta y
-(2\pi\alpha')
\cG_{\rm A}^{u N}\delta a_{N}
\right)\right]_{z=-\infty}^{z=+\infty}\ ,
\\
 \delta S_{\rm CS}
&=&\mbox{(EOM)}+\frac{N}{8\pi}\int d^2 x\,
\left[a_+\delta a_--a_-\delta a_+\right]_{z=-\infty}^{z=+\infty}\ ,
\\
 \delta S_{a\pm}
&=&-\frac{N}{8\pi}\int_{z\ra\pm\infty}\hspace{-2ex} d^2 x\,
(a_+\delta a_-+a_-\delta a_+)\ ,
\end{eqnarray}
where $\mbox{(EOM)}$ denotes the bulk terms that give the
equations of motion \req{yEOM} and \req{aEOM}, while
$\cG$, $\cG_{\rm S}^{MN}$ and $\cG_{\rm A}^{MN}$ are defined as
in equations \req{cG}, \req{MS} and \req{MA}.

In Appendix \ref{eomDBI-CS},
it is shown that the gauge field $a_M$ satisfying the
equations of motion can always be decomposed as 
\begin{eqnarray}
a_M=a_M^{(0)}+b_M\ ,
\end{eqnarray}
where $a_M^{(0)}$ is a flat connection satisfying (\ref{flatcon}),
and $b_M$ is defined by
\begin{eqnarray}
 b_\pm&\equiv&
 \mp \frac{4\pi}{N}(2\pi\alpha')T_{\rm 3d}
\sqrt{-\cG}\,\cG_A^{u\mp}\ ,
\\
 b_u&\equiv&
  \frac{4\pi}{N}(2\pi\alpha')T_{\rm 3d}
\sqrt{-\cG}\,\cG_A^{+-}\ .
\end{eqnarray}
Therefore, for the on-shell configurations, the variation becomes
\begin{eqnarray}
\delta S_{\rm total}
&=&
-T_{\rm 3d}\int d^2 x\,
\left[
\sqrt{-\cG}
\frac{u^2}{R^2}\cG_S^{u N}\del_N y\,\delta y
\right]_{z=-\infty}^{z=+\infty}
\nn\\
&&
-\frac{N}{4\pi}\int_{z\ra+\infty}\hspace{-2ex} d^2 x\,
(a^{(0)}_-\delta a_+ +b_+\delta a_-)
-\frac{N}{4\pi}\int_{z\ra-\infty}\hspace{-2ex} d^2 x\,
(a^{(0)}_+\delta a_- +b_-\delta a_+)
\ .
\label{surf}
\end{eqnarray}
For our solution (\ref{asymy})--(\ref{asymam2}), we have
\begin{eqnarray}
&&\sqrt{-{\cG}}
\frac{u^2}{R^2}\cG_S^{u N}\del_N y= c_y\ ,
\\
&&b_\pm(z)=\frac{c_\pm}{8\pi\alpha'}f^{(\mp)}(z)\sim\cO(u^{-4})\ ,
~~(z\ra\pm\infty)\ .
\label{bp}
\end{eqnarray}
Then, the boundary conditions
(\ref{abdry}) and (\ref{ybdry}) imply that the surface terms in
(\ref{surf}) with $\delta y$ and $b_{\pm}\delta a_\mp$ vanish,
because $\cO(u^0)$ terms in $\delta y$, and
 $\cO(u^4)$ terms in $\delta a_\mp$ ($z\ra\pm\infty$)
are zero when the sources are fixed.
In order to make sure that the surface terms in (\ref{surf}) with
$a_\mp^{(0)}\delta a_\pm$ vanish, we impose a boundary condition as
$a_\mp^{(0)}\sim\cO(u^0)$ at $z\ra\pm\infty$.

\subsection{Gauge invariance and anomaly }
\label{gaugeinvanom}

One may wonder the consistency of the counterterm (\ref{Saa}), because
it is not gauge invariant. In fact the counterterm (\ref{Saa}) is
needed to ensure the gauge invariance.
Let us clarify this point.
Under the gauge transformation
\begin{eqnarray}
a\ra a+d\lambda\ , 
\label{gaugetrans}
\end{eqnarray}
$S_{\rm CS}$ and $S_{a\pm}$ transform as
\begin{eqnarray}
\delta_\lambda S_{\rm CS}
&=&-\frac{N}{8\pi}\left(
\int_{z\ra+\infty}\hspace{-2ex}d^2x\,  \lambda f_{+-}
-\int_{z\ra-\infty}\hspace{-2ex}d^2x\,  \lambda f_{+-}
\right)\ ,
\label{deltaScs}
\\
\delta_\lambda S_{a\pm}
&=&
\frac{N}{8\pi}
\int_{z\ra\pm\infty}d^2x\,\left(\lambda\del_+a_-+\lambda\del_-a_+\right)\ .
\end{eqnarray}
Here, we have dropped the surface terms at $|x^\pm|\ra\infty$.
Then, assuming that all the other counterterms are gauge invariant,
the total action transforms as
\begin{eqnarray}
 \delta_\lambda S_{\rm total}
=\frac{N}{4\pi}\left(
\int_{z\ra+\infty}\hspace{-2ex}d^2x\,  \lambda \del_-a_+
+\int_{z\ra-\infty}\hspace{-2ex}d^2x\,  \lambda \del_+a_-
\right)\ .
\label{anom1}
\end{eqnarray}
Because of the boundary condition (\ref{abdry}),
$a_\pm$ do not diverge at $z\ra\pm\infty$, and
the total action is invariant under the gauge transformation with
$\lambda$ vanishing at $z\ra\pm\infty$. 
Note that for general field configuration with the boundary condition
(\ref{abdry}), (\ref{deltaScs}) is non-vanishing. The gauge invariance
is guaranteed only after the counterterms $S_{a\pm}$ are added.

When the
$U(1)$ symmetry associated to the current
$J_-$ is gauged, 
the gauge transformation of this $U(1)$ symmetry
\begin{eqnarray}
 \cA_+\ra \cA_+ +\del_+ \Lambda\ ,
\label{cAgauge}
\end{eqnarray}
is realized by imposing a boundary condition for $\lambda$ as
\begin{eqnarray}
 \lambda(x^a,z)\ra\Lambda(x^a)\ ,~~~(z\ra +\infty)\ .
\end{eqnarray}
As we have seen in (\ref{anom1}), the D7-brane action is not invariant
under this gauge transformation and transforms as
\begin{eqnarray}
 \delta_\lambda S_{\rm total}
=\frac{N}{4\pi}
\int_{z\ra +\infty}\hspace{-2ex}d^2x\,  \Lambda \del_-\cA_+
\ .
\label{anom2}
\end{eqnarray}
This expression precisely agrees with the anomalous transformation of
the generating function for correlation functions in the
dual field theory induced by one loop diagrams of the chiral fermion on
the defect.
In fact, omitting the supergravity action,
the on-shell value of the action $S_{\rm total}$ is identified as
\begin{eqnarray}
 e^{iS_{\rm total}^{\rm o.s.}(\cA)}
\propto \int\cD\psi\cD A\, e^{iS_{\rm 3d}(\psi,A,\cA)}\ ,
\end{eqnarray}
where $S_{\rm 3d}(\psi,A,\cA)$ is the action of the 3-dimensional
$SU(N)$ YM-CS theory with defect given by the sum of
(\ref{SA}) (with $k=1$, $k'=0$), (\ref{Spsi}) and (\ref{AJ}).
Then, the gauge transformation (\ref{cAgauge}) of this equation
and (\ref{anom2}) imply the anomaly equation%
\footnote{
See \cite{Jensen:2010em,Yee:2011yn}
and section \ref{anomsbem} for closely related derivations.
}
\begin{eqnarray}
\del_+\VEV{J_-} = 
-2\p_+\VEV{J^+} = 
\frac{N}{2\pi}\del_-\cA_+\ ,
\label{anomeq}
\end{eqnarray}
reproducing equation~\req{anomaly} for the case $k-k'=1$.

\subsection{Correlation functions}

Comparing the asymptotic behavior (\ref{asymy})--(\ref{asymam2})
of the solution to the boundary conditions (\ref{abdry})--(\ref{ybdry}),
the sources in our configuration are identified as
\begin{eqnarray}
 \cA_\pm =a_\pm^{(0)}|_{z\ra\pm\infty}\ ,~~~
\cC_\mp=\frac{c_\mp}{8\pi\alpha'}\frac{R^8}{u_*^4}e^\xi\ ,~~~
\cY^{(\pm)}=\pm L\ .
\end{eqnarray}
The correlation functions can be computed by differentiating the
on-shell action (\ref{ostotal}) with respect to these sources.
Using the expressions (\ref{surf})--(\ref{bp}),
the one point functions for the operators at $y=+L$ ($z\ra+\infty$)
are obtained as
\begin{eqnarray}
\VEV{\cO_y}&=&\frac{\delta S_{\rm total}^{\rm o.s.}}{\delta\cY^{(+)}}
=-T_{\rm 3d} c_y
=-\frac{N}{8\pi(2\pi\alpha')^2}\frac{u_*^3}{R^2}
\sqrt{1+\frac{c_+c_-}{u_*^2}}\ ,
\label{VEVOy}
\\
\VEV{J^+}&=&\frac{\delta S_{\rm total}^{\rm o.s.}}{\delta\cA_+}
=-\frac{N}{4\pi}a^{(0)}_-|_{z\ra +\infty}\ ,
\label{VEVJm}
\\
\VEV{\cO_+}&=&\frac{\delta S_{\rm total}^{\rm o.s.}}{\delta\cC_-}
=-\frac{N}{4\pi}\frac{c_+}{8\pi\alpha'}\frac{u_*^4}{R^8} e^{-\xi}\ .
\label{VEVOp}
\end{eqnarray}

The two (or higher) point functions can be obtained by differentiating
these expressions with respect to the sources.

\subsubsection{Condensation of $\cO_y$}

\begin{figure}[ht]
\begin{center}
\begin{picture}(210,130)(0,0)
\put(10,0){\includegraphics[scale=0.7]{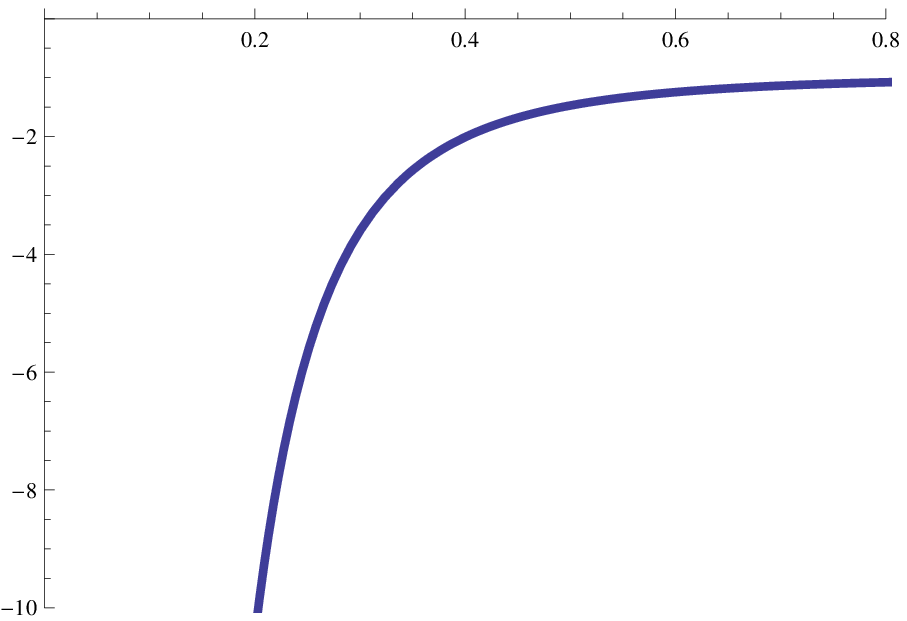}}
\put(-5,50){\makebox(0,0){$\VEV{\cO_y}$}}
\put(200,123){\makebox(0,0){$L$}}
\put(10,123){\makebox(0,0){$0$}}
\end{picture}
\parbox{80ex}{
\caption{$\VEV{\cO_y}|_{\cC_\pm=0}$ as a function of $L$, normalized by
 its absolute value at $L\ra\infty$.}
\label{FigOy}}
\end{center}
\end{figure}

In particular, (\ref{VEVOy}) implies that
$\VEV{\cO_y}$ is non-zero even when
the external sources $\cA_\pm$ and $\cC_\pm$ are turned off.
When $\cC_\pm=0$, as depicted in Fig.~\ref{FigOy},
the absolute value $|\VEV{\cO_y}|$ is a monotonically
decreasing function of $L$ and the asymptotic value is
\begin{eqnarray}
\VEV{\cO_y}|_{\cC_\pm=0,L\ra\infty} 
=-\frac{N}{8\pi(2\pi\alpha')^2}\frac{u_0^3}{R^2}
=-\frac{N \lambda_{\rm 3d} M_{\rm KK}^2}{64\pi^2}\ ,
\end{eqnarray}
where $\lambda_{\rm 3d}\equiv g_{\rm 3d}^2 N=g_s M_{\rm KK}N$
is the 't Hooft coupling.

For small $L$ ($u_*\gg u_0$), the equation (\ref{L})
is approximated as
\begin{eqnarray}
L \simeq 2f_0
\frac{R^2}{u_*}\ ,
\label{smallL}
\end{eqnarray}
where
\begin{eqnarray}
f_0=\frac{\sqrt{\pi}\Gamma(2/3)}{2\Gamma(1/6)}
\simeq 0.216\ .
\label{f0}
\end{eqnarray}
Then, we obtain
\begin{eqnarray}
\VEV{\cO_y}|_{\cC_\pm=0} 
\simeq
-\frac{N f_0^3}{\pi^2}\frac{ \lambda_{\rm 3d}}{M_{\rm KK}}\frac{1}{L^3}
\end{eqnarray}
for small $L$. The $L$ dependence is consistent with the
conformal symmetry at UV.

\subsubsection{Anomaly, symmetry breaking and edge modes}
\label{anomsbem}

The one point function for the current (\ref{VEVJm})
was obtained and analyzed in closely related systems in
\cite{Jensen:2010em,Yee:2011yn}. Let us comment
on some of the interesting consequences obtained
by following the arguments given in these papers.

Recall that $a^{(0)}$ is a flat connection and it can be
written as
\begin{eqnarray}
 a^{(0)}=d\varphi\ ,
\label{varphi}
\end{eqnarray}
with a real scalar field $\varphi$. Then, (\ref{VEVJm}) and the
analogous equation for $J^-$ on the other defect
placed at $y=-L$ ($z\ra-\infty$) can be
written as
\begin{eqnarray}
\VEV{J^\pm}=-\frac{N}{4\pi}\del_\mp\varphi^{(\pm)}\ ,
\label{Jvarphi}
\end{eqnarray}
where we have defined
\begin{eqnarray}
 \varphi^{(\pm)}\equiv \lim_{z\ra\pm\infty}\varphi\ .
\end{eqnarray}
As we have argued in section \ref{gaugeinvanom}, the gauge
transformation (\ref{gaugetrans}) acts trivially
at $z\ra\pm\infty$, and therefore
these $\varphi^{(\pm)}$ cannot be gauged away.
Because $a_\pm^{(0)}|_{z\ra\pm\infty}=\cA_\pm$,
they are related to the external fields $\cA_\pm$ as
\begin{eqnarray}
 \cA_\pm=\del_\pm\varphi^{(\pm)}\ .
\label{Avarphi}
\end{eqnarray}
Then, it is easy to see that (\ref{Jvarphi}) reproduces the anomaly
equation (\ref{anomeq}).

When $\cA_\pm=0$, (\ref{Avarphi}) implies that
$\varphi^{(+)}$ and $\varphi^{(-)}$ are chiral and anti-chiral boson
which depend only on $x^-$ and $x^+$, respectively.
These modes correspond to the gapless edge modes that exist at the
boundary of FQH states, which are described by the CS theory.
As the relation (\ref{Jvarphi}) suggests, they are related to the chiral
(anti-chiral) fermions on the defects by bosonization.
The equation (\ref{Jvarphi}) also suggests that $\varphi^{(\pm)}$
are the Nambu-Goldstone modes associated with the spontaneous breaking
of the $U(1)\times U(1)$ symmetry generated by the currents $J_\pm$.
Actually, the vacuum expectation value of $\varphi^{(+)}+\varphi^{(-)}$
is unphysical, since it can be shifted by a
 constant shift $\varphi\ra \varphi+(\mbox{constant})$,
which is the redundancy of the definition of $\varphi$
in (\ref{varphi}). Therefore, the diagonal subgroup $U(1)_{\rm diag}$
of the $U(1)\times U(1)$ is unbroken.
On the other hand, the other combination
\begin{eqnarray}
 \varphi^{(+)}- \varphi^{(-)}=\int_{-\infty}^\infty dz\,(a_z-b_z)\ ,
\end{eqnarray}
is unambiguously defined and it corresponds to the Nambu-Goldstone (NG)
mode associated with the symmetry breaking
 $U(1)\times U(1)\ra U(1)_{\rm diag}$.
This is analogous to the chiral symmetry breaking in holographic QCD
as discussed in \cite{Sakai:2004cn} and more directly related
to the 2-dimensional version studied in \cite{Yee:2011yn}.
Note that this NG mode lives in 2-dimension, which is
justified only in the large $N$ limit.
When $N$ is finite, the quantum corrections for the holographic
description should be taken into account and
the symmetry will be restored as shown in
\cite{Mermin:1966fe,Coleman:1973ci}.\footnote{For a holographic version
of this statement, see \cite{Anninos:2010sq}.}

\subsubsection{Correlations between the two defects}
\label{twopt}

\begin{figure}[ht]
\begin{center}
\begin{picture}(260,130)(0,0)
\put(20,0){\includegraphics[scale=.9]{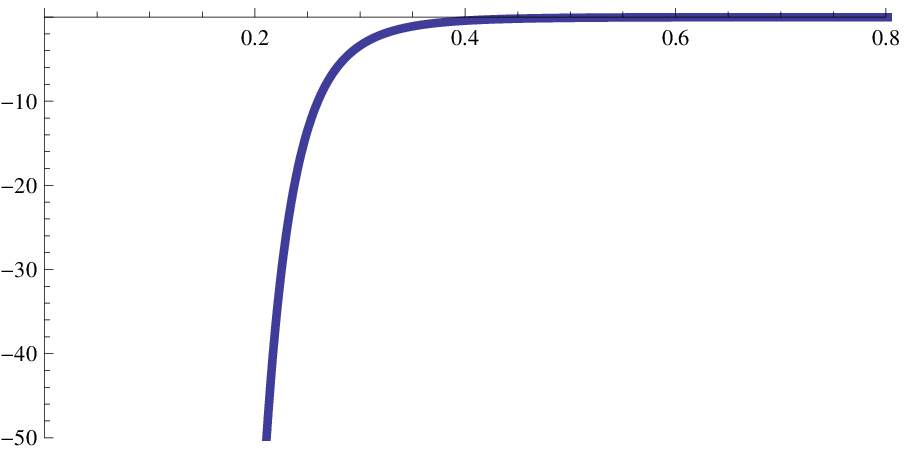}}
\put(-5,50){\makebox(0,0){$-u_*^8 e^{-2\xi}$}}
\put(260,113){\makebox(0,0){$L$}}
\put(22,113){\makebox(0,0){$0$}}
\end{picture}
\parbox{80ex}{
\caption{$\VEV{\cO_-\cO_+}$ 
as a function of $L$.}
\label{FigOpOm}}
\end{center}
\end{figure}

Let us consider the two point function $\VEV{\cO_-\cO_+}$ at vanishing source
$\cC_{\pm}=0$, where $\cO_-$ and $\cO_+$ are dimension 5 operators
placed on the defect at $y=-L$ and $y=+L$, respectively.
Differentiating (\ref{VEVOp}) with respect to the constant source
$\cC_+$, we obtain
\begin{eqnarray}\label{eq551}
\VEV{\cO_-\cO_+}\equiv
\int d^2x'
 \VEV{\cO_-(x)\cO_+(x')}|_{\cC_\pm=0}
=-\frac{N}{4\pi}\frac{u_*^8}{R^{16}} e^{-2\xi}\Big|_{\cC_\pm=0}\ .
\end{eqnarray}
The behavior of this two point function as a function of $L$ is depicted
in Fig.~\ref{FigOpOm}.

For large $L$ ($u_*\ra u_0$), we can show
\begin{eqnarray}
\xi\,|_{\cC_\pm=0} &=& 4\int_{u_*}^\infty du
\left(\sqrt{\frac{u^4-u_0^4}{u^6-u_*^6}}-\frac{1}{u}
\right)+\frac{4 u_0^4}{R^2 u_*^3} L
\nn\\
&\simeq&
-c_0+2L M_{\rm KK}\ ,
\end{eqnarray}
where
\begin{eqnarray}
 c_0\equiv -4\int_1^\infty dt\left(
\sqrt{\frac{t^4-1}{t^6-1}}-\frac{1}{t}\right)\simeq 0.260\ .
\end{eqnarray}
and hence the two point function behaves as
\begin{eqnarray}
\VEV{\cO_-\cO_+}
\simeq -\frac{N}{4\pi}\frac{u_0^8\,e^{2c_0}}{R^{16}} e^{-(2M_{\rm KK})2L}
= -\frac{N}{1024\pi} M_{\rm KK}^8 e^{2c_0} e^{-(2M_{\rm KK})2L}
\label{VEV55}
\end{eqnarray}
for large $L$.
This behavior suggests that the lightest particle that couples to
$\cO_+$ and $\cO_-$ has mass $2M_{\rm KK}$.

For small $L$ ($u_*\gg u_0$), using (\ref{smallL}), we get
\begin{eqnarray}
\xi\,|_{\cC_\pm=0}
\simeq
\frac{4}{3}\log 2+\frac{M_{\rm KK}^4}{32f_0^3}L^4
\ ,
\end{eqnarray}
and
\begin{eqnarray}
\VEV{\cO_-\cO_+}
\simeq
-\frac{N2^{10/3}f_0^8}{\pi}\frac{1}{L^{8}}\ .
\end{eqnarray}
This falloff is consistent with conformal scaling for a position-space two-point function of the dimension five operators ${\cal O}_\pm$ due to the additional integral in the definition \eqref{eq551}.

\section{Free energy, phase transition, and confinement}
\label{freeenergy}

In this section, we consider the free energy of our system
at zero temperature%
\footnote{
The results in this section are valid for $T<T_c$, where $T_c$ is the
critical temperature given in (\ref{Tc}). See section \ref{finT} for a
discussion of the case $T>T_c$.}
using the holographic description.
We are mainly interested in the
$L$ dependence of the free energy and study the phase structure
by varying the positions of the defects.
Here, we set $\cA_\pm=\cC_\mp=0$.

\subsection{Free energy}
\label{freeenergy1}

\begin{figure}[ht]
\begin{center}
\begin{picture}(200,200)(0,0)
\put(5,10){\includegraphics[scale=0.7]{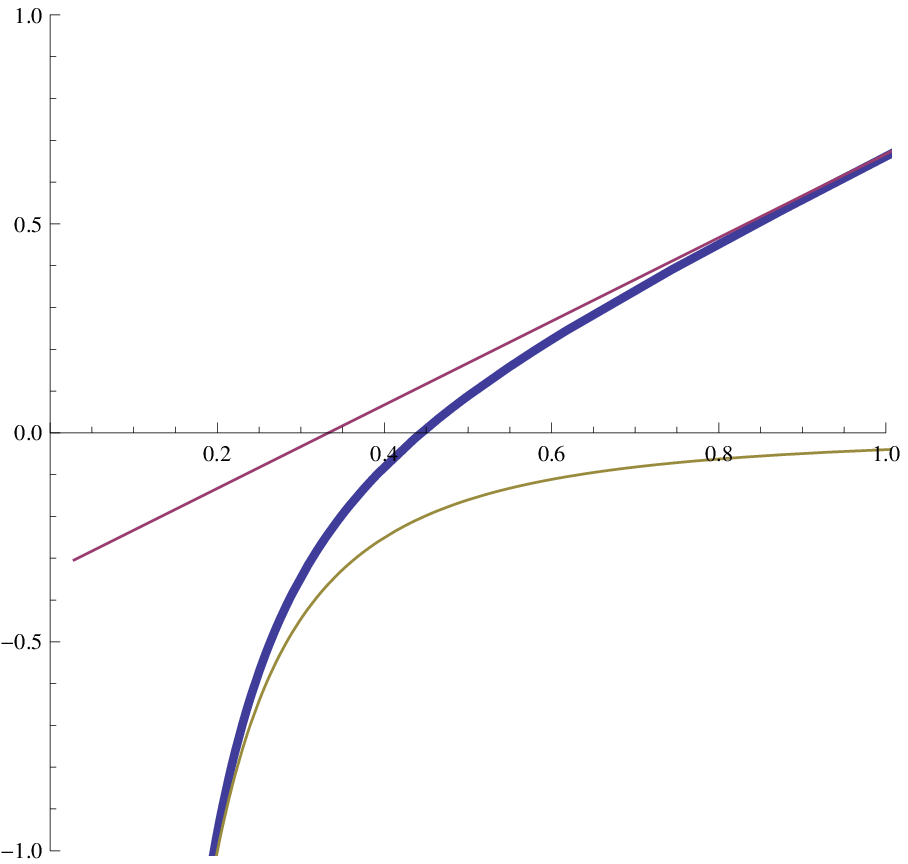}}
\put(175,108){\makebox(0,0){$\frac{u_0}{R^2}L$}}
\put(0,168){\makebox(0,0){$\frac{R}{u_0^2}\cF$}}
\put(45,114){\makebox(0,0){\footnotesize $\frac{u_0}{R^2}L-a_0$}}
\put(37,105){\vector(1,-3){6}}
\put(130,54){\makebox(0,0){\footnotesize
$-4f_0^3\left(\frac{u_0}{R^2}L\right)^{-2}$}}
\put(120,65){\vector(-1,3){6}}
\end{picture}
\parbox{70ex}{
\caption{Free energy as a function of $L$.
}
\label{FFtil}}
\end{center}
\end{figure}

Following the standard dictionary of holography, the free energy for our
configuration, neglecting the $L$-independent part, is proportional to
the on-shell action (\ref{ostotal}). We define a function $\cF(L)$
proportional to the free energy by
\begin{eqnarray}
 S^{\rm o.s.}_{\rm total}|_{\cA_\pm=\cC_\mp=0}
\equiv -2T_{\rm 3d}\int d^2x\, \cF(L)\ .
\label{SandF}
\end{eqnarray}
Setting $a_\pm^{(0)}=c_\pm=0$ in (\ref{ostotal}), we obtain
\begin{eqnarray}
 \cF(L)
&=&
\frac{u_*^2}{2R}
\left(
\int_{-\infty}^{+\infty}dz
\left(\frac{(1+z^2)^3}{\wt F(z)}
-|z|\right)-1
\right)\Bigg|_{c_\pm=0}
\nn\\
&=&
\int_{u_*}^{\infty}du\,
\frac{u}{R}
\left(\frac{u^5}{\sqrt{(u^4-u_0^4)(u^6-u_*^6)}}-1
\right)
-\frac{u_*^2}{2R}\ ,
\label{FL0}
\end{eqnarray}
where $u_*$ is related to $L$ by (\ref{L}).
A plot of $\cF(L)$ is depicted in Fig.~\ref{FFtil}.

For $u_*\gg u_0$ (small $L$), $\cF$ and $L$ behave as
\begin{eqnarray}
\cF \simeq -f_0\frac{u_*^2}{R}\ ,
~~~L \simeq 2f_0
\frac{R^2}{u_*}\ ,
\label{expFL}
\end{eqnarray}
where $f_0$ is defined in (\ref{f0}),
and hence we obtain
\begin{eqnarray}
 \cF(L)
\simeq
-4f_0^3\frac{R^3}{L^2}
\label{FL}
\end{eqnarray}
for small $L$.

For $u_*\ra u_0$ (large $L$), we have
\begin{eqnarray}
 \cF(L)\simeq \frac{u_0^2}{R}
\left(
\frac{u_0}{R^2}L-
a_0
\right)
\label{linear}
\end{eqnarray}
with
\begin{eqnarray}
a_0=\half-\int_1^\infty dt\left(
\sqrt{\frac{t^6-1}{t^4-1}}-t
\right)
\simeq 0.333
\ .
\end{eqnarray}
To get this, note that (\ref{FL0}) can be written as
\begin{eqnarray}
\cF&=&
\frac{1}{R}\int_{u_*}^{\infty}du
\left(
\frac{u^6-u_*^6}{\sqrt{(u^4-u_0^4)(u^6-u_*^6)}}-u
+\frac{u_*^6}{\sqrt{(u^4-u_0^4)(u^6-u_*^6)}}
\right)
-\frac{u_*^2}{2R}
\nn\\
&=&
\frac{1}{R}\int_{u_*}^{\infty}du
\left(
\sqrt{\frac{u^6-u_*^6}{u^4-u_0^4}}-u
\right)
+\frac{u_*^3}{R^3}L
-\frac{u_*^2}{2R}\ .
\label{eq:F-derivation}
\end{eqnarray}
Then, (\ref{linear}) can be easily obtained by taking $u_*\ra u_0$.

The linear behavior of the leading term in
(\ref{linear}) is analogous to the linear
potential for a quark - anti-quark pair in confining gauge theories.
Instead of inserting a quark - anti-quark pair, we have considered
a defect - anti-defect pair and observed similar linear behavior.
In fact, they have the same geometric origin in the holographic
description. In the case of the quark - anti-quark potential,
the linear behavior is due to the fact that the string tension
is non-zero at the minimum value of the radial coordinate
$u$.\cite{Witten:1998zw} In our case, the string is replaced with the
probe D7 brane and the linear behavior in (\ref{linear})
is understood from the fact that the D7-brane tension evaluated at
$u=u_0$ is non-zero,
which is evident from the geometry.
In fact, the D7-brane tension at $u=u_0$ is given by
\begin{eqnarray}
 T_{\rm 3d}\sqrt{-g_{00}g_{11}g_{yy}}\,\Big|_{u=u_*}
=T_{\rm 3d}\frac{u_0^3}{R^3}\ ,
\end{eqnarray}
and the factor $u_0^3/R^3$ agrees with
the coefficient of $L$ in the leading term of (\ref{linear})
for large $L$.

Note that the problem of finding D7-brane configurations
and the on-shell values of the D7-brane action (for $a_M=0$)
is mathematically equivalent to the holographic computation 
of entanglement entropy when $\tau$ is interpreted as time
after double Wick rotation.
This is because the dilaton field is constant in our background
and the D7-brane configurations are given
by minimal surfaces with given boundary conditions.
Since the D7-brane action is proportional to the area of the
D7-brane world-volume, the on-shell value of the action
gives the area of the minimal surface, which is proportional to
the entanglement entropy as proposed in \cite{Ryu:2006bv,Ryu:2006ef}.%
Therefore, the free energy $\cF(L)$ is proportional to the entanglement
entropy between the regions $|y|<L$ and $|y|>L$ up to a divergent $L$
independent constant.
In fact, the entanglement entropy for the AdS soliton background has been 
studied in \cite{Nishioka:2006gr,Klebanov:2007ws,Ben-Ami:2014gsa}
and many of the formulas and figures shown below (section \ref{finT})
agree with those appearing in these papers.

\subsection{Phase transition}

\begin{figure}[ht]
\begin{center}
\begin{picture}(200,110)(0,0)
\put(5,0){\includegraphics[scale=0.7]{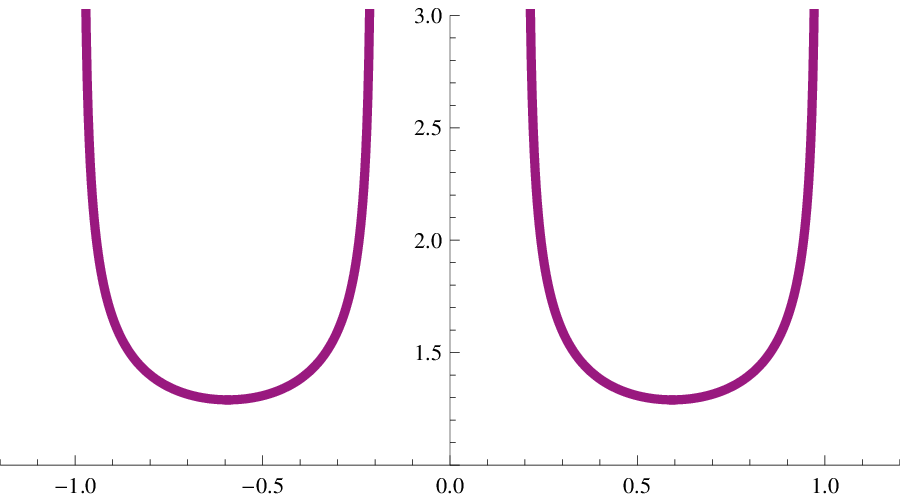}}
\put(179,16){\makebox(0,0){$\frac{u_0}{R^2}y$}}
\put(96,108){\makebox(0,0){$\frac{u}{u_0}$}}
\put(98,95){\vector(-1,0){1}}
\put(96,95){\vector(1,0){14}}
\put(103,94){\line(2,-3){20}}
\put(126,61){\makebox(0,0){$l$}}
\put(22,106){\makebox(0,0){\footnotesize (1)}}
\put(80,106){\makebox(0,0){\footnotesize (2)}}
\put(112,106){\makebox(0,0){\footnotesize (3)}}
\put(170,106){\makebox(0,0){\footnotesize (4)}}
\end{picture}
~~
\begin{picture}(200,110)(0,0)
\put(5,0){\includegraphics[scale=0.7]{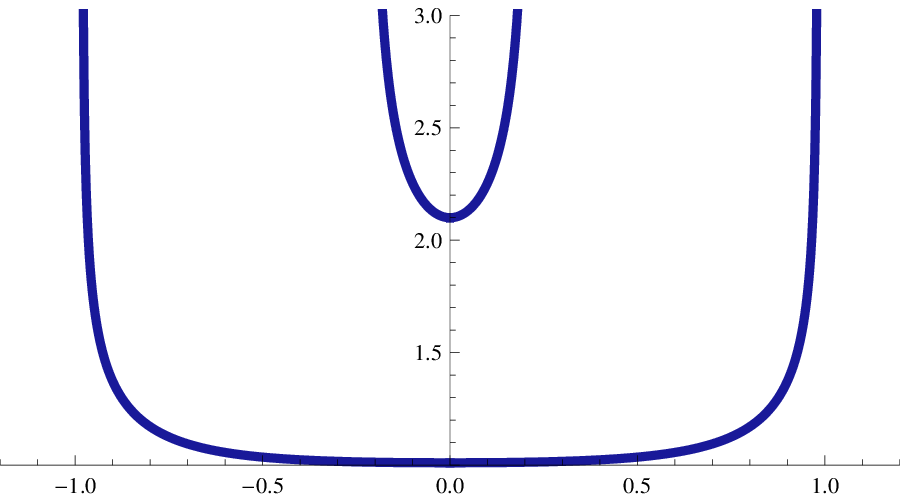}}
\put(179,16){\makebox(0,0){$\frac{u_0}{R^2}y$}}
\put(96,108){\makebox(0,0){$\frac{u}{u_0}$}}
\put(98,95){\vector(-1,0){1}}
\put(96,95){\vector(1,0){12}}
\put(102,95){\line(1,-2){15}}
\put(121,61){\makebox(0,0){$l$}}
\put(22,106){\makebox(0,0){\footnotesize (1)}}
\put(82,106){\makebox(0,0){\footnotesize (2)}}
\put(110,106){\makebox(0,0){\footnotesize (3)}}
\put(170,106){\makebox(0,0){\footnotesize (4)}}
\end{picture}
\parbox{70ex}{
\caption{$UU$-phase (left) and $\breve U$-phase (right).
}
\label{phase}}
\end{center}
\end{figure}

\begin{figure}[ht]
\begin{center}
\begin{picture}(200,180)(0,0)
\put(5,10){\includegraphics[scale=0.7]{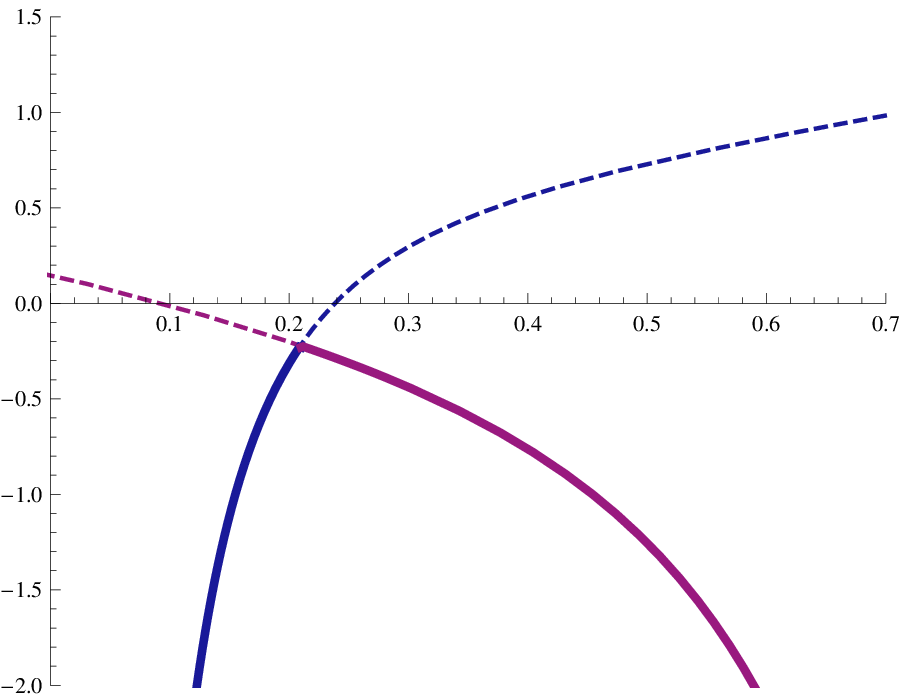}}
\put(180,97){\makebox(0,0){$l$}}
\put(0,155){\makebox(0,0){$\frac{R}{u_0^2}\cF$}}
\put(85,32){\makebox(0,0){\footnotesize $\breve U$-phase}}
\put(65,35){\vector(-1,1){10}}
\put(155,67){\makebox(0,0){\footnotesize$UU$-phase}}
\put(145,60){\vector(-1,-1){10}}
\end{picture}
\parbox{70ex}{
\caption{Free energy of $\breve U$-phase and $UU$-phase.
}
\label{phase2}}
\end{center}
\end{figure}

If there are more than one components of U-shaped D7 branes,
phase transitions occur by changing the parameters of the system.
As a simple example, consider placing four defects (1)$\sim$(4)
at
(1) $y=-L$, (2) $y=-l$, (3) $y=+l$, (4) $y=+L$ with $0<l<L$ such that
the CS level $(-k)$ for the $SU(N)$ YM-CS theory is
$k=1$ for $l<|y|<L$, and $k=0$ for $|y|<l$ and $L<|y|$.
The holographic dual of this system contains
two U-shaped D7 branes as in Fig.~\ref{phase}.
There are two solutions with the same boundary conditions.
We call the left and right sides of Fig.~\ref{phase} the
$UU$-phase and the $\breve U$-phase, respectively.
When the parameter $l$ is smaller (larger) than a critical value $l_c$,
the $\breve U$-phase ($UU$-phase) is favored.
The free energy of these configurations is depicted in Fig.~\ref{phase2}.
In terms of the two point function $\VEV{\cO_+\cO_-}$ discussed in
section \ref{twopt}, there are correlations between defects (1) and (2),
and also between (3) and (4) for $l>l_c$:
\begin{eqnarray}\label{VEV616}
\VEV{\cO_-^{(1)}\cO_+^{(2)}} \ne 0\ ,~~~
\VEV{\cO_-^{(3)}\cO_+^{(4)}} \ne 0\ .~~~
(l>l_c)
\end{eqnarray}
As $l$ decreases and the defect (2) and (3) approach,
there is a phase transition at critical value of $l=l_c$
and the $\breve U$-phase is favored for $l<l_c$. Then, in this phase,
the correlated pairs are changed to
\begin{eqnarray}
\VEV{\cO_+^{(2)}\cO_-^{(3)}} \ne 0\ ,~~~
\VEV{\cO_-^{(1)}\cO_+^{(4)}} \ne 0\ .~~~
(l<l_c)
\end{eqnarray}
It is interesting that the correlation between the farthest pair (1) and
(4) appears when $l$ is small.

\subsection{Confinement}
\label{sec:confinement}

Pure YM in 3-dimensions is known to be confining at a scale of order
$\Lambda_{\rm YM}\sim g_{\rm 3d}^{2}$, giving rise to a mass gap 
$m_{\rm gap}\sim\Lambda_{\rm YM}$.
Pure CS theory, on the other hand, does not confine: it is a
topological field theory, whose expectation values compute topological  
invariants of the spacetime manifold~\cite{Witten:1988hf}.
In YM-CS theory, the CS term induces a tree-level mass for gluons,
$m_{CS} = \frac{|k_{CS}|g_{\rm 3d}^2}{2\pi}$, and the topological gap competes
with the confining behavior of the YM action.
It is a non-trivial question which behavior will dominate in the infrared.

To determine which is realized in our system we should compute the expectation
value of a Wilson loop along a contour $\cC$ in some representation $\cR$,
$\VEV{\Tr_\cR\cP \exp\left(i\oint_\cC A\right)}$.
The contour most often used consists of a rectangle with length $T$ in the
temporal direction and width $W$ in a spatial direction, with $T\gg W$.
If large loops have an expectation value $e^{i\cW_\cC}$ with $\cW_\cC$ 
proportional to minus the area $WT$, then the theory is confining.
This is the famous area law.
On the other hand, if the behavior is topological then for large loops,
the expectation value will be finite and independent of the loop's size 
and shape (up to local counterterms).

In the holographic context it is practical to make the computation instead
in Euclidean time, with metric
\be
ds_E^2 = \frac{u^2}{R^2}(dt_E^2+dx^2+dy^2+f(u)d\tau^2) 
+ \frac{R^2}{u^2}\frac{du^2}{f(u)} + R^2 d\Omega_5^2 \ .
\label{eq:euclidean}
\ee
As usual we have identified $t_E \sim t_E + \beta$, with $\beta$ the inverse
temperature. 
It is important here that we take the temperature to be much smaller than 
the compactification scale $M_{\rm KK}$.

Having compactified the time direction, we will consider a pair of Wilson
lines wrapping the Euclidean time direction, with opposite orientation and 
at fixed separation $W$. 
Our discussion here will be restricted to the case where the level $(-k)$ is the
same everywhere, and there are no defects, corresponding to $k$ D7 branes
located at the tip ($u=u_0$) of the AdS soliton.

We start by reviewing the case $k=0$, with no D7 branes
 at the soliton tip.\cite{Witten:1998zw,Brandhuber:1998er} 
Wilson lines are computed in the semi-classical limit by the holographically 
renormalized Euclidean worldsheet 
action of a string which attaches to the Wilson line at the asymptotic 
boundary,
\be
S^{\rm euc.}_\mathrm{string} = 
\frac{1}{2\pi\alpha'}\int_{\Sigma} d^2\sigma \, \sqrt{\det(g_{ab})} 
+ \mbox{(counterterms)} \ 
\ee
where $g_{ab}$ is the pullback to the worldsheet of the spacetime 
metric~\req{eq:euclidean}. 
The loop we are interested in is invariant under time translations, 
so the shape of the worldsheet is determined by the profile in the
$y$-$u$ plane, $y(u)$.
With this ansatz the Nambu-Goto action takes the form
\be
S^{\rm euc.}_{\rm NG}
 = \frac{1}{2\pi\alpha'}\int_\Sigma dt_E\ du \sqrt{\Delta(u)}
\qquad\quad
\Delta(u) = \frac{1}{f(u)} + \frac{u^4}{R^4}\!\left(\frac{dy}{du}\right)^2\ ,
\ee
resulting in the equation of motion
\begin{equation}
\frac{u^4}{R^4}\frac{y'(u)}{\sqrt{\Delta}} = c
\end{equation}
with $c$ a constant. The solution is
\begin{equation}
y(u) = y_0 + \int_{u_*}^u 
	\frac{cR^4\, d\hat u}{\sqrt{(\hat u^4-u_0^4)(\hat u^4-u_*^4)}}
\end{equation}
where $c = \pm u_*^2/R^2$, from which we find the distance between 
the endpoints 
\begin{equation}
W = 2\int_{u_*}^{\infty}\frac{u_*^2R^2\ du}{\sqrt{(u^4-u_0^4)(u^4-u_*^4)}}\ .
\end{equation}
As with the D7-brane configuration, the Wilson line at constant $\tau$
corresponds to $u_* \ge u_0$.

As usual, the on-shell action is divergent, but can be regularized by 
cutting off the ambient spacetime along the cutoff surface $u=u_\Lambda$.
Using the relation $\sqrt{\Delta} = \frac{u^4}{cR^4}\frac{dy}{du}$, the
NG action takes the form
\begin{equation}
S_{\rm NG} = \frac{2\beta}{2\pi\alpha'}\int_{u_*}^{u_\Lambda}
	\frac{u^4\ du}{\sqrt{(u^4-u_0^4)(u^4-u_*^4)}} 
\simeq \frac{2u_{\Lambda}\beta}{2\pi\alpha'}\ .
\end{equation}
To renormalize the action we must include the counterterm
\begin{equation}
S_{\rm ct} = -\frac{R}{2\pi\alpha'}\int_{\p\Sigma}dt\sqrt{\gamma}
= -\frac{1}{2\pi\alpha'}\left(\int_{\rm left}+\int_{\rm right}\right) 
	u_\Lambda\, dt
\end{equation}
with $\gamma$ the pullback of the Euclidean AdS soliton metric to the 
intersection of the worldsheet with the cutoff surface $u=u_\Lambda$.

The renormalized action is 
$S_{\rm ren} = \lim_{u_\Lambda\to\infty}(S_{\rm NG}+S_{\rm ct})$.
It is convenient to introduce the free 
energy $\cF$ associated with the Wilson line,
$S_{\rm ren}=\beta\cF$.
The free energy then takes the form
\begin{equation}
\cF = \frac{1}{\pi\alpha'}\left[ \int_{u_*}^\infty du \left(
\frac{u^4}{\sqrt{(u^4-u_0^4)(u^4-u_*^4)}}-1\right) - u_* \right] .
\end{equation}
The behavior of $\cF$ for $W\gg M_{\rm KK}^{-1}$ can be obtained using the same
method as \req{eq:F-derivation}, giving the asymptotic $W$-dependence 
\begin{equation}
\cF \simeq
\frac{u_0^2}{2\pi\alpha'R^2}W - \frac{u_0}{\pi\alpha'} +
\cO(e^{-M_{\rm KK}W})\ .
\end{equation}
Thus for sufficiently large $W$ we find the area law expected in a 
confining theory.

\begin{figure}[ht]
\begin{center}
\begin{picture}(289,181)(0,0)
\put(5,5){\includegraphics{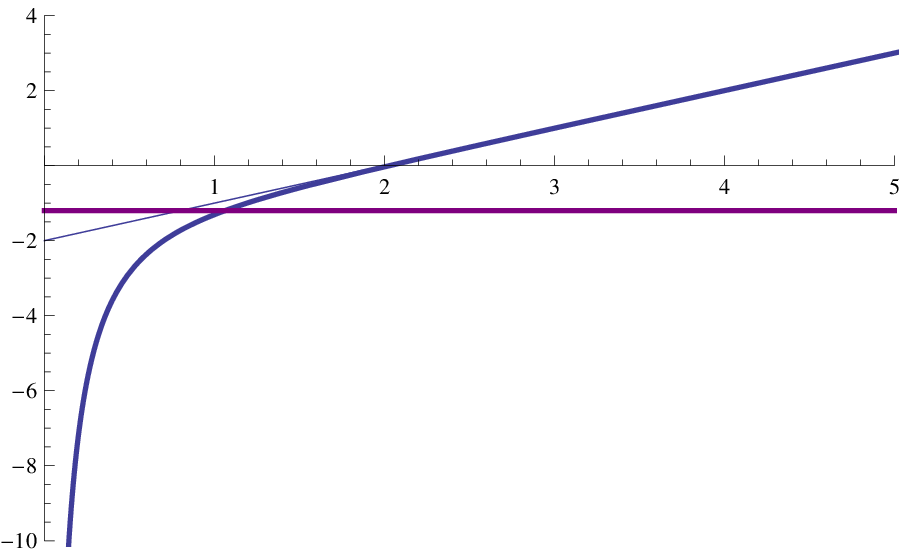}}
\put(265,130){\makebox(0,0){$\frac{u_0}{R^2}W$}}
\put(37,160){\makebox(0,0){$\frac{2\pi\alpha'}{u_0}\cF$}}
\put(100,145){\makebox(0,0){\footnotesize connected phase}}
\put(138,138){\vector(1,-1){10}}
\put(215,85){\makebox(0,0){\footnotesize disconnected phase}}
\put(172,90){\vector(-1,1){10}}
\end{picture}
\parbox{70ex}{
\caption{Free energy $\cF$ of the Wilson line anti-line pair as a
 function of separation $W$.}
\label{fig:wilson}}
\end{center}
\end{figure}

The computation changes qualitatively when the CS level $(-k)$ of
the boundary is non-zero, because in this case there are $k$ D7 branes 
located at the soliton tip on which the worldsheet can end.
Now there is a competing configuration in which two 
disconnected worldsheets stretch between the loops on the boundary 
and the branes at the soliton tip.
In the semi-classical limit, we can ignore backreaction from both the 
gravitational sector and the gauge fields on the brane, in which case the
preferred configuration is $y={\rm constant}$.
The renormalized worldsheet action then takes the form
\begin{equation}
S_{\rm ren} = \frac{2\beta}{2\pi\alpha'}
	\left[ \int_{u_0}^\infty du \left( \frac{1}{\sqrt{f(u)}} - 1 \right)
		- u_0 \right]\ .
\end{equation}
The resulting free energy is a constant $\cF=2\cF_0$, with
\begin{equation}
\cF_0 = - \frac{u_0}{2\pi\alpha'}\frac{\sqrt{\pi}\Gamma(3/4)}{\Gamma(1/4)}
= -\frac{\Gamma(3/4)}{2\ \Gamma(1/4)}\sqrt{\lambda_{\rm 3d}M_{\rm KK}}
\simeq -0.1690 \times \sqrt{\lambda_{\rm 3d} M_{\rm KK}}\ .
\label{eq:F0}
\end{equation}
(Recall that $\lambda_{\rm 3d}=Ng_sM_{\rm KK}$.)

The comparison of the free energy in the two phases is shown in 
figure~\ref{fig:wilson}.
We see that, for $W>W_{\rm crit}\simeq \frac{R^2}{u_0}\times 1.063$,
the phase with the worldsheet ending on the D7 branes has lower free energy, 
indicating a first order transition from the connected phase (which would
show an area law at large separation) to the disconnected
phase that shows a perimeter law.

Note that if we interpret the Wilson line as the insertion of a heavy quark,
the free energy $\cF_0$ corresponds to a self-energy.
When computing Wilson line expectation values it is natural to choose a 
renormalization scheme in which the perimeter law contributions vanish 
precisely, which can be accomplished by adding the finite local counterterm
$S_{\rm ct}^{(2)} = -\int_{\p\Sigma} \cF_0$.

With this modification, the computation of large Wilson lines in the
$k\ne 0$ phase reduces to the computation of correlators
the Wilson lines in CS theory on the D7 brane, in agreement with the
claim of~\cite{Fujita:2009kw}
that this system provides an explicit realization of level-rank duality.
We conclude that in the semi-classical regime, and with $|k|\ll N$, the theory
is in a topological phase and does not confine.

\subsection{Chiral condensate}
\label{sec:cc}

\begin{figure}[ht]
\begin{center}
\begin{picture}(310,130)(0,0)
\put(5,10){\includegraphics[scale=0.8]{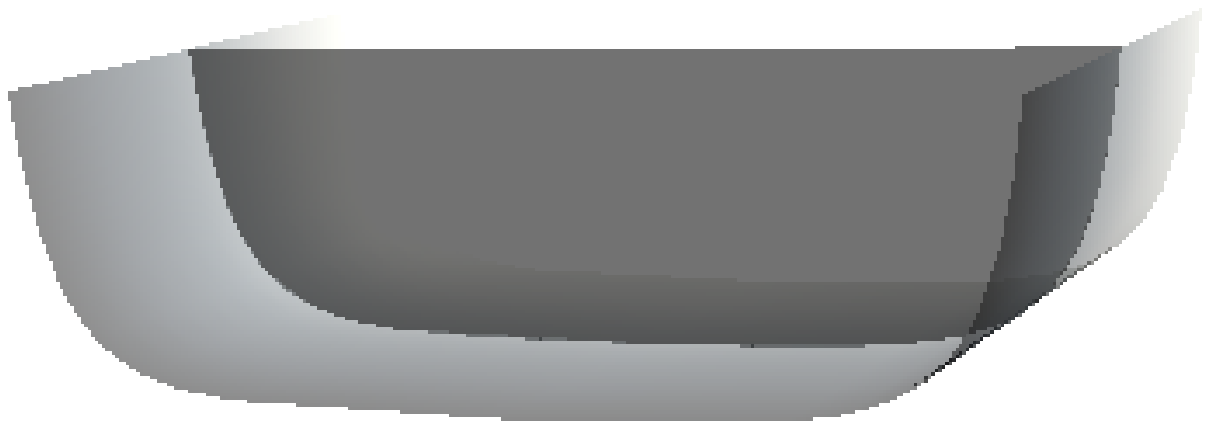}}
\put(0,10){\makebox(0,0){D7 brane}}
\put(5,15){\vector(1,1){22}}
\put(150,112){\makebox(0,0){string worldsheet}}
\put(270,10){\vector(1,0){30}}
\put(270,10){\vector(0,1){30}}
\put(270,10){\vector(1,1){20}}
\put(305,10){\makebox(0,0){$y$}}
\put(270,45){\makebox(0,0){$u$}}
\put(302,37){\makebox(0,0){$(t,x)$}}
\end{picture}
\parbox{70ex}{
\caption{Open string configuration dual to the open Wilson line ending on
two parallel level-changing defects.\label{fig:open-string-config}}
}
\end{center}
\end{figure}

In the presence of a defect--anti-defect pair, 
we expect a chiral condensate to form between the chiral fermions living on
the two defects at zero temperature. 
The chiral condensate in question takes the form
$\langle \psi_L^\dagger \cP e^{i\int_R^L A} \psi_R\rangle$, where gauge
invariance forces us to include an open Wilson 
line stretching between the fermion insertions on the two defects.
The holographic dual of this object is the open string worldsheet that 
attaches on the AdS soliton boundary
to the Wilson line.\cite{Aharony:2008an,Hashimoto:2008sr}
The dual configuration is shown in Fig.~\ref{fig:open-string-config}.

In the semi-classical limit, the expectation value takes the form
$\langle \psi_L^\dagger \cP e^{i\int_R^L A} \psi_R\rangle = e^{-S_{\rm ren}}$,
with $S_{\rm ren}$ the renormalized Euclidean worldsheet action as
derived in the previous section.
For the present configuration, it takes the form
\begin{equation}
S_{\rm ren} = \frac{1}{\pi\alpha'}\left[
\int_{u_*}^{\infty}du\left(\frac{y(u)}{\sqrt{f(u)}} - L \right) - Lu_*
\right]  ,
\label{eq:cc-S}
\end{equation}
where $y(u)$ is as in~\req{eq:y} (here we set $y_0=0$ and $u_{\rm min}=u_*$).

For large $L$, it is convenient to introduce the object 
\begin{equation}
d(u) = y(u) - L = -R^2u_*^3\int_u^\infty 
\frac{d\hat u}{\sqrt{(\hat u^4-u_0^4)(\hat u^6-u_*^6)}}\ ,
\end{equation}
in which case we may write
\begin{equation}
S_{\rm ren} = \frac{1}{\pi\alpha'}
	\int_{u_*}^\infty du \frac{d(u)}{\sqrt{f(u)}}
+ \frac{(2L)}{2\pi\alpha'}
	\left[\int_{u_*}^\infty
	 du\left(\frac{1}{\sqrt{f(u)}}-1\right)-u_*\right]\ . 
\label{eq:cc}
\end{equation} 
When $L\gg M_{\rm KK}^{-1}$, $u_*$ approaches $u_0$ and
the second term of \req{eq:cc} depends linearly on $L$,
taking the form $2L\cF_0$ (with $\cF_0$ the free energy \req{eq:F0} of an
isolated Wilson line). 
It is instructive to consider the dependence of the first term on $L$,
which in the limit of large $L$ contributes a constant to the free energy:
\be
\lim_{L\to\infty}(S_{\rm ren}-2L\cF_0) =
\frac{1}{\pi\alpha'}\int_{u_0}^\infty du\frac{d(u)|_{u_*=u_0}}{\sqrt{f(u)}}  
= \frac{R^2}{\pi\alpha'} J_0
\ee
with
\be
J_0 = -\int_1^\infty dx \frac{x^2}{\sqrt{x^4-1}}\int_x^\infty
\frac{dv}{\sqrt{(v^4-1)(v^6-1)}}
\simeq -0.299\ .
\ee
This should be understood as (twice) the contribution due to an isolated 
endpoint of an infinitely extended open Wilson line. 
Therefore we may write
\be
S_{\rm ren} = (2L) \cF_0 + \frac{R^2}{\pi\alpha'}J_0 
+ I(L)
\label{eq:large-L-OWL-action}
\ee
where the remainder $I(L)=\cO(e^{-\sqrt{6}M_{\rm KK}L})$ decays
exponentially to zero as $L\to\infty$.

\begin{figure}[ht]
\begin{center}
\begin{picture}(298,181)(0,0)
\put(5,5){\includegraphics{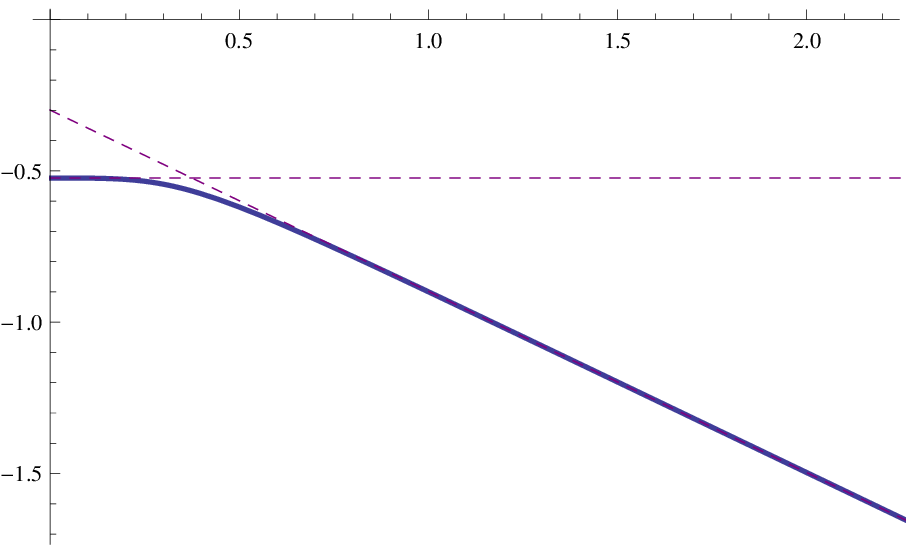}}
\put(45,15){\makebox(0,0){$\frac{\pi\alpha'}{R^2}S_{\rm ren}$}}
\put(265,145){\makebox(0,0){$\frac{u_0}{R^2}L$}}
\put(80,135){\makebox(0,0){$\frac{2\pi\alpha'\cF_0}{R^2}L+J_0$}}
\put(55,130){\vector(-1,-1){10}}
\put(220,127){\makebox(0,0){$\frac{\pi\alpha'}{R^2}S_0$}}
\put(205,122){\vector(-1,-1){10}}
\end{picture}
\parbox{70ex}{
\caption{Plot of the open Wilson line action $S_{\rm ren}$ (in units of
 $R^2/\pi\alpha'$) as a function of length.} 
\label{fig:open-wilson}}
\end{center}
\end{figure}

For $L\ll M_{\rm KK}^{-1}$, $d(u)$ can be approximated by a hypergeometric
function
\be
d(u) \simeq -\frac{R^2}{4u_*}\left(\frac{u_*}{u}\right)^4
{}_2F_1\bigl(\frac{1}{2},\frac{2}{3};\frac{5}{3};\frac{u_*^6}{u^6}\bigr)\ .
\ee
Using with the asymptotic behavior \req{smallL} of $L$, we find that for 
$L\ll M_{\rm KK}^{-1}$,
\be
S_{\rm ren} \simeq 
S_0 = \frac{1}{\pi\alpha'}\left[\int_{u_*}^\infty du\,
d(u) - u_*L\right]_{u_0=0}
= -\frac{R^2}{\pi\alpha'}\cdot\frac{\pi}{6}
= -\frac{\sqrt{4\pi g_sN}}{6}\ .
\ee
The action $S_{\rm ren}$ for general values of $L$ is shown in figure 
\ref{fig:open-wilson}.

Note that for large separations, the chiral condensate in the
semiclassical limit is $e^{-S_{\rm ren}}\sim e^{2L\cdot |\cF_0|}$. 
The exponential growth with length of the correlation function is surprising,
as one might expect it rather to decay exponentially at a rate 
determined by the scale $M_{\rm KK}$.
In our case, we can see that the exponential dependence on $L$ arises
because of the self energy of the Wilson line derived in
section~\ref{sec:confinement}.
These results are analogous to the behavior of the chiral condensate for 
D8-$\overline{\rm D8}$ defects in the D4 brane worldvolume theory 
discussed in~\cite{Aharony:2008an}, which also found a
similar exponential
dependence on separation as the endpoints of the chiral condensate operator
were given a large separation parallel to the defects.
In particular, they find that at strong coupling, the dominant contribution
to the chiral condensate operator comes from the Wilson line, rather than
the fermion bilinear.

When defining the renormalized Wilson line operator, we have the option of 
including a finite counterterm of the form $S_{\rm ct} = a\int ds$, which
is sufficient to eliminate the linear behavior at large $L$ of 
eq.~\req{eq:large-L-OWL-action}. 
Similarly, we may insert a constant counterterm at the string endpoints,
allowing us to eliminate the $J_0$ contribution. 
This suggests that the quantity that is physically relevant to the 
computation of the expectation value of the chiral condensate itself 
is the function $I(L)$
of~\req{eq:large-L-OWL-action}.%
\footnote{See \cite{McNees:2008km} for related discussion in holographic QCD.}

\section{Finite temperature}
\label{finT}

\subsection{Background metric and D7-brane configuration}

In order to introduce finite temperature $T$, we compactify
the Wick rotated time $t_E\equiv ix^0$ as
\begin{eqnarray}
 t_E\sim t_E+\beta
\end{eqnarray}
with inverse temperature $\beta=1/T$.
It is known that there is a phase transition at the critical
temperature
\begin{eqnarray}
 T_c\equiv \frac{M_{\rm KK}}{2\pi}
=\frac{u_0}{R^2\pi}\ ,
\label{Tc}
\end{eqnarray}
corresponding to the confinement/deconfinement
transition.~\cite{Witten:1998zw,Kruczenski:2003uq}
The background metric for the low temperature phase $T<T_c$
is the same as (\ref{metric}). For the high temperature phase $T>T_c$,
it is changed to
\begin{eqnarray}
 ds^2=\frac{u^2}{R^2}\left(f_T(u)dt_E^2+dx^2+dy^2+d\tau^2
\right)+\frac{R^2}{u^2}\frac{du^2}{f_T(u)}
+R^2 d\Omega_5^2\ ,
\label{metricT}
\end{eqnarray}
where $x=x^1$, $y=x^2$, $\tau=x^3$ and
\begin{eqnarray}
 f_T(u)=1-\frac{u_T^4}{u^4}
\end{eqnarray}
with
\begin{eqnarray}
 u_T=\pi R^2 T\ .
\end{eqnarray}
Note that $T>T_c$ implies $u_T>u_0$.

\begin{figure}[ht]
\begin{center}
\begin{picture}(200,140)(0,0)
\put(5,5){\includegraphics[scale=0.7]{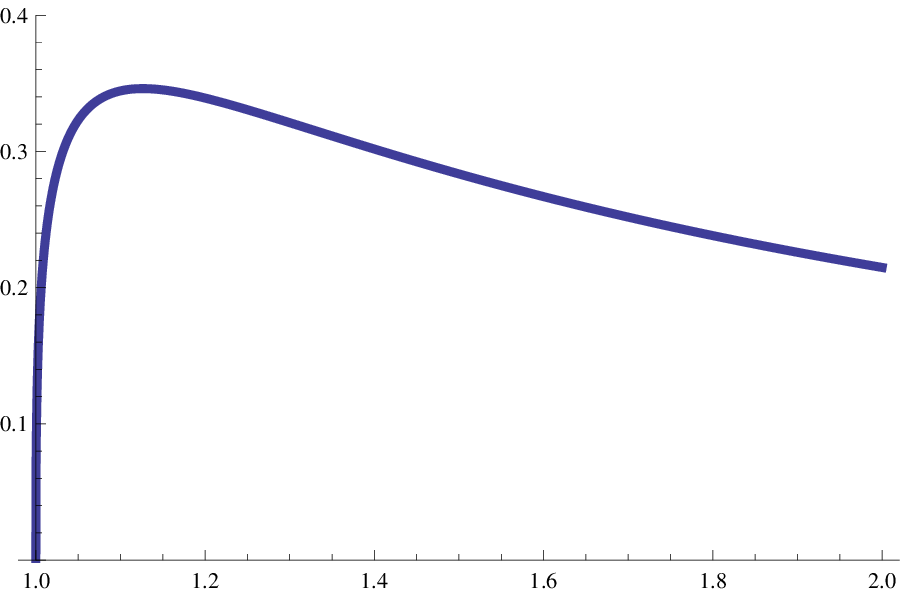}}
\put(195,15){\makebox(0,0){$\frac{u_*}{u_T}$}}
\put(0,130){\makebox(0,0){$\frac{u_T}{R^2}L$}}
\put(13,107){\line(1,0){20}}
\put(3,107){\vector(1,0){8}}
\put(-13,107){\makebox(0,0){\footnotesize $0.346$}}
\put(34,12){\line(0,1){95}}
\put(34,3){\vector(0,1){8}}
\put(35,-2){\makebox(0,0){\footnotesize $1.13$}}
\end{picture}
\parbox{70ex}{
\caption{$L$ as a function of $u_*$.
}
\label{figL}}
\end{center}
\end{figure}

The U-shaped D7-brane configuration for $T>T_c$ 
with $f_{MN}=0$ and $\tau=0$ is given by
\begin{eqnarray}
 y(u)=R^2\int_{u_*}^u
\frac{du'}{\sqrt{(u'^4-u_T^4)
\left(\frac{u'^2(u'^4-u_T^4)}{u_*^2(u_*^4-u_T^4)}-1\right)}}\ .
\label{finTUsol}
\end{eqnarray}
(See Appendix \ref{solT} for details.)
A plot of $L\equiv\lim_{u\ra\infty}|y(u)|$
as a function of $u_*$ is shown in Fig.~\ref{figL}.
As one can see from Fig.~\ref{figL}, there is a maximum value of $L$
around
\begin{eqnarray}
L_{\rm max}\simeq 0.346 \times\frac{R^2}{u_T}\ ,
\end{eqnarray}
for the U-shaped solution to exist.
For $L<L_{\rm max}$,
there are two solutions with the same $L$.

\begin{figure}[ht]
\begin{center}
\begin{picture}(200,120)(0,0)
\put(5,0){\includegraphics[scale=0.7]{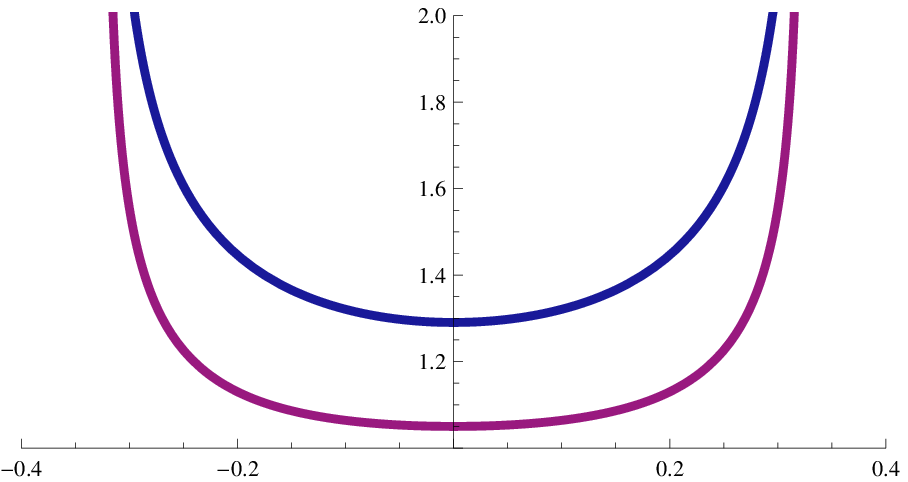}}
\put(185,18){\makebox(0,0){$\frac{u_T}{R^2}y$}}
\put(100,106){\makebox(0,0){$u/u_T$}}
\end{picture}
~~
\begin{picture}(200,120)(0,0)
\put(5,0){\includegraphics[scale=0.7]{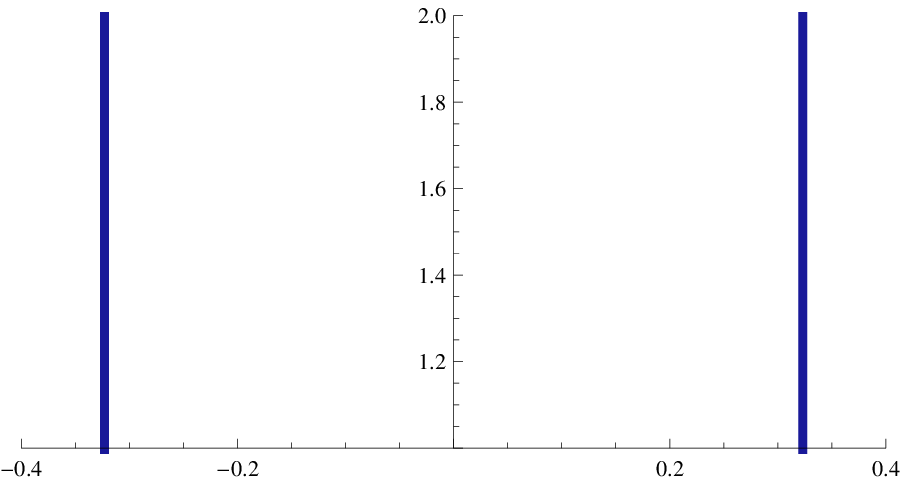}}
\put(185,18){\makebox(0,0){$\frac{u_T}{R^2}y$}}
\put(100,106){\makebox(0,0){$u/u_T$}}
\end{picture}
\parbox{70ex}{
\caption{U-shaped solutions (left) and a
disconnected solution (right).
There are two U-shaped solutions with the same $L$ as shown in the left
 figure. }
\label{finTconfig}}
\end{center}
\end{figure}

There is another type of solution given by $y={\rm constant}$.
In this case, the D7 brane and $\ol{\rm D7}$ brane are disconnected
and placed at $y=L$ and $y=-L$, respectively. They cover the entire
$(t_E,x,u)$ directions without any singularities.
Unlike the U-shaped solution considered above, the disconnected
solutions exist for all $L$. 
These solutions are shown in Fig.~\ref{finTconfig}.

\subsection{Free energy and phase transition}
For the U-shaped solution (\ref{finTUsol}), the function $\cF$ defined
in (\ref{SandF}) is given by
\begin{eqnarray}
 \cF=\int_{u_*}^\infty du\,
\frac{u}{R}\left(
\frac{1}{\sqrt{1-\frac{u_*^2(u_*^4-u_T^4)}{u^2(u^4-u_T^4)}}}-1
\right)-\frac{u_*^2}{2R}\ .
\end{eqnarray}

For the disconnected solution $y={\rm constant}$, we get
\begin{eqnarray}
 \cF=-\frac{u_T^2}{2R}\ ,
\end{eqnarray}
which is independent of $u_*$.

For $u_*\gg u_T$ (small $L$), we have
\begin{eqnarray}
 L\simeq 2f_0\frac{R^2}{u_*}\ ,
~~~\cF\simeq -f_0\frac{u_*^2}{R}\ ,
\end{eqnarray}
and
\begin{eqnarray}
\cF\simeq -4 f_0^3\frac{R^3}{L^2}\ ,
\end{eqnarray}
which are the same as (\ref{expFL})
and (\ref{FL}). This is expected
because the asymptotic behavior in the region $u\gg u_T$
is not affected by the temperature.

\begin{figure}[ht]
\begin{center}
\begin{picture}(200,200)(0,0)
\put(5,0){\includegraphics[scale=0.7]{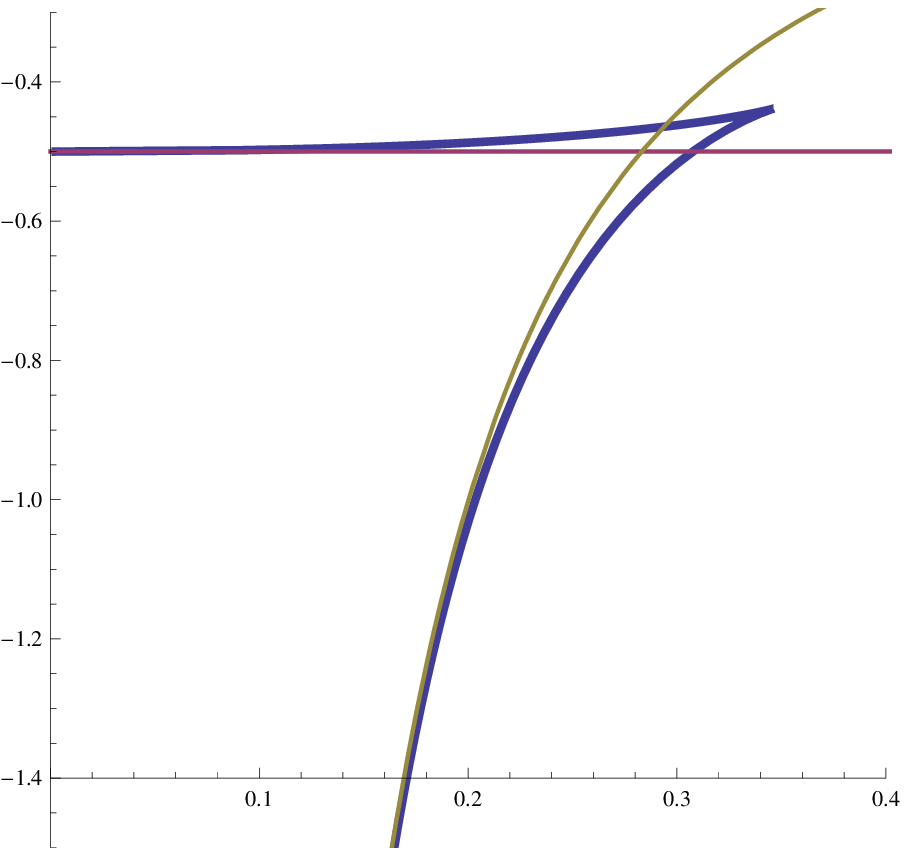}}
\put(200,15){\makebox(0,0){$\frac{u_T}{R^2}L$}}
\put(0,173){\makebox(0,0){$\frac{R}{u_T^2}\cF$}}
\put(102,170){\makebox(0,0){\footnotesize
$-4f_0^3\left(\frac{u_T}{R^2}L\right)^{-2}$}}
\put(128,162){\vector(3,-1){20}}
\put(26,130){\vector(-1,1){10}}
\put(42,124){\makebox(0,0){\footnotesize $\frac{u_*}{u_T}=1$}}
\put(162,125){\vector(0,4){25}}
\put(170,117){\makebox(0,0){\footnotesize $\frac{u_*}{u_T}\simeq 1.13$}}
\put(90,10){\vector(-1,-4){5}}
\put(110,0){\makebox(0,0){\footnotesize $\frac{u_*}{u_T}\ra\infty$}}
\put(6,141){\vector(1,0){8}}
\put(-8,140){\makebox(0,0){\footnotesize $-0.5$}}
\put(175,175){\vector(0,-1){32}}
\put(173,183){\makebox(0,0){disconnected}}
\put(120,75){\vector(-1,1){10}}
\put(146,68){\makebox(0,0){U-shaped}}
\end{picture}
\parbox{70ex}{
\caption{$\cF$ as a function of $L$.
}
\label{finTF}}
\end{center}
\end{figure}

Another configuration with small $L$ is obtained when
$u_*$ approaches $u_T$. In the limit $u_*\ra u_T$, we have
\begin{eqnarray}
 L&\simeq&-\frac{R^2}{4u_T}
\sqrt{\epsilon_*}\log\epsilon_*\ ,
\\
\cF&\simeq&-\frac{u_T^2}{2R}\left(
1+\frac{\epsilon_*}{4}\log\epsilon_*\right)\ ,
\end{eqnarray}
where $\epsilon_*\equiv (u_*/u_T)^4-1$.
\begin{figure}[ht]
\begin{center}
\begin{picture}(200,140)(0,0)
\put(5,5){\includegraphics[scale=0.7]{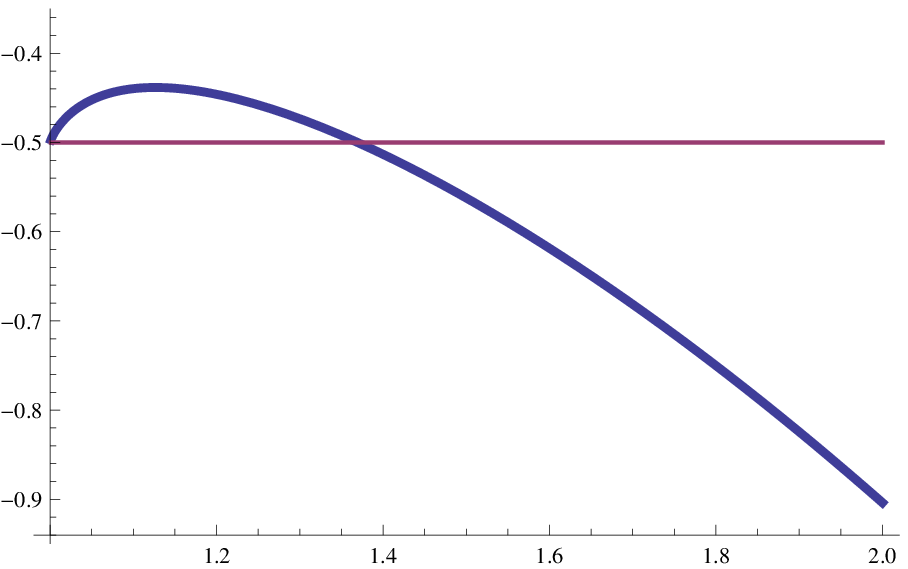}}
\put(195,15){\makebox(0,0){$\frac{u_*}{u_T}$}}
\put(5,125){\makebox(0,0){$\frac{R}{u_T^2}\cF$}}
\put(16,103){\line(1,0){20}}
\put(6,103){\vector(1,0){8}}
\put(-13,103){\makebox(0,0){\footnotesize $-0.438$}}
\put(37,12){\line(0,1){90}}
\put(37,3){\vector(0,1){8}}
\put(35,-2){\makebox(0,0){\footnotesize $1.13$}}
\put(78,12){\line(0,1){80}}
\put(78,3){\vector(0,1){8}}
\put(78,-2){\makebox(0,0){\footnotesize $1.37$}}
\put(150,105){\vector(0,-1){10}}
\put(150,113){\makebox(0,0){disconnected}}
\put(150,65){\vector(-1,-1){10}}
\put(176,68){\makebox(0,0){U-shaped}}
\end{picture}
\parbox{70ex}{
\caption{$\cF$ as a function of $u_*$.
}
\label{Fus}}
\end{center}
\end{figure}
The behavior of $\cF$ as a function of $u_*$ and $L$ are shown
in Fig.~\ref{Fus} and Fig.~\ref{finTF}, respectively.

Fig.~\ref{figL} and Fig.~\ref{Fus} suggest that
both $L$ and $\cF$ take maximum values at $u_*/u_T\sim 1.13$.
In fact, one can show a relation
\begin{eqnarray}
 \frac{\del\cF(u_*)}{\del u_*}=\frac{u_*^3}{R^3}\sqrt{1-
\frac{u_T^4}{u_*^4}}\, \frac{\del L(u_*)}{\del u_*}\ ,
\end{eqnarray}
which implies that $L$ and $\cF$ take maximum at the same point.

\begin{figure}[ht]
\begin{center}
\begin{picture}(200,200)(0,0)
\put(5,0){\includegraphics[scale=0.7]{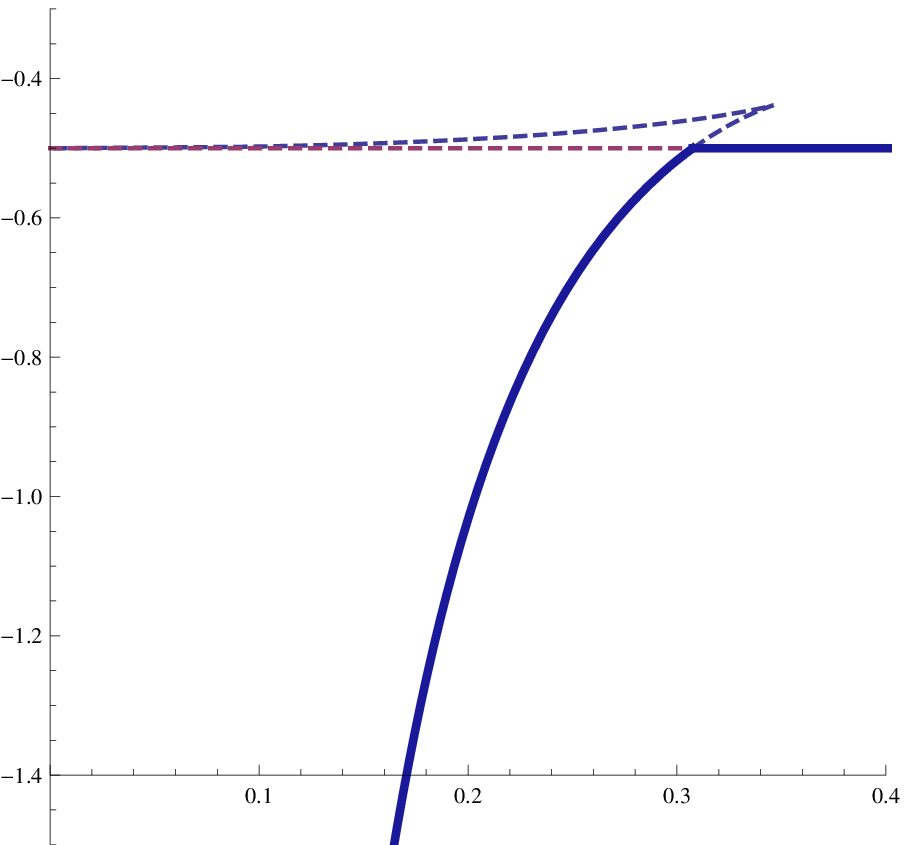}}
\put(200,15){\makebox(0,0){$\frac{u_T}{R^2}L$}}
\put(0,173){\makebox(0,0){$\frac{R}{u_T^2}\cF$}}
\put(146,15){\line(0,1){125}}
\put(146,6){\vector(0,1){8}}
\put(146,0){\makebox(0,0){\footnotesize $0.308$}}
\end{picture}
\parbox{70ex}{
\caption{Free energy as a function of $L$.
}
\label{finTF2}}
\end{center}
\end{figure}

Therefore, there is a critical value of $L$ around
\begin{eqnarray}
 L_c\simeq \frac{R^2}{u_T}\times 0.308\ ,
\end{eqnarray}
at which the brane configuration jumps:
\begin{eqnarray}
\begin{array}{ccl}
L<L_c&\Rightarrow&\mbox{U-shaped solution}\,, \\
L>L_c&\Rightarrow&\mbox{disconnected solution}\,.
\end{array}
\end{eqnarray}
A plot of the minimum values of $\cF$
is shown in Fig.~\ref{finTF2}.

This phenomenon is similar to the behavior of
the probe D8 brane discussed in \cite{Aharony:2006da}
in the context of the holographic QCD based on D4/D8-brane
system.\cite{Sakai:2004cn}
In the phase described by the disconnected solution,
the $U(1)\times U(1)$ symmetry, which is broken to
$U(1)_{\rm diag}$ at $T=0$ as discussed in section \ref{anomsbem},
is restored. This is because the two boundaries are disconnected
and $\varphi^{(+)}$ and $\varphi^{(-)}$ can be shifted independently,
unlike the case for the U-shaped configuration
discussed in section \ref{anomsbem},

\newpage

\section{Summary and discussion}
\label{summary}

This work dealt with level-changing defects in YM-CS field theory, 
as realized holographically within the construction of~\cite{Fujita:2009kw}. 
We found explicit solutions for the probe brane profiles dual to these defects,
providing a clear geometric understanding of their behavior under level-rank 
duality.
After holographic renormalization, we computed the zero-momentum 
correlation functions for operators transforming trivially under the 
(ultraviolet) $SO(6)$ $R$-symmetry.
Our analysis shows that the system exhibits several interesting phenomena 
including anomalies and (in the limit of infinite $N$) the spontaneous breaking 
of global symmetries localized on the defects. 
Systems with multiple defects furthermore exhibit interesting phase
transitions in which operators localized on defect pairs become
correlated or uncorrelated, depending on the relative separations of the
defects.
In the finite temperature case, we find that this phase transition has an
interesting structure as the temperature rises above the critical temperature
for the ($k=0$) confinement-deconfinement phase transition.

As we argued in section \ref{anomsbem}, the gapless edge mode
found in the 3-dimensional $U(1)$ DBI-CS theory on the probe D7 brane with
two boundaries corresponds to the Nambu-Goldstone mode associated
with the chiral symmetry breaking (an analog of the pion) in large $N$
2-dimensional QCD with one massless flavor. This observation suggests
interesting relations between the physics of the FQHE and 2-dimensional
QCD. In fact, there is a direct correspondence between these two
seemingly unrelated theories, because both of them are
governed by $U(1)$ CS theory at low energies: the effective theory
of mesons in 2-dimensional QCD is given by 3-dimensional DBI-CS theory
on a D7 brane \cite{Yee:2011yn}, while the $U(1)$ CS theory (for the
statistical gauge field) is 
an effective theory of the Laughlin states of the FQHE.  
The particle that couples to the statistical $U(1)$ gauge field
with the unit charge is the quasiparticle (or quasihole) of the FQH state,
and should correspond to the end point of a fundamental
string attached to the D7 brane; 
in 2-dimensional QCD, this is interpreted as an external quark. 
Since the CS level is $N$, the quasiparticle
carries an electric charge $1/N$, corresponding to
the baryon number charge of the quark. 
Therefore, the electron (an object with unit electric charge) in FQH 
state corresponds to the baryon in 2-dimensional QCD.
It would be interesting to investigate this correspondence in more
detail.

We offered further evidence that in the IR limit the model becomes
non-abelian CS theory with level-changing defects, and thus
resembles (the non-Abelian generalization of) the FQHE in the presence
of defects (or edges). Not only does the IR
theory exhibit a gap in the  
bulk between the defects, the Wilson loop evaluated in the bulk between
the defects exhibits the topological (perimeter law) behavior
expected of a CS theory when the CS level is non-vanishing, and
confining (area law) behavior expected of pure YM theory when the CS
level vanishes.  
This suggests a number of interesting further questions. 
The first regards the Hall response.
This was computed in \cite{Fujita:2009kw} in the absence of defects, 
both in field theory and its holographic dual.
However, the physical Hall current should actually be carried by the
edge modes, being localized on the defects.
It would be interesting to verify that this edge current is correctly
reproduced by our system in the presence of a background electric field. 
Another is how flux attachment, recently discussed in two different
holographic setups in
\cite{Jokela:2013hta,Jokela:2014wsa,Lippert:2014jma}, is realized in
the setup considered here. 

Condensed matter physicists have discussed a variety of
experimental setups that 
can probe the charge and statistics of the gapless quasiparticle
excitations at the edge of FQH samples.  
In the simplest setup, an electric voltage applied between the two edges
of a FQH sample leads,  
at zero temperature, to tunneling of quasiparticle excitations between
the edges.  
Assuming that the edges are described by 1-dimensional Luttinger liquids
with Luttinger exponent $g$, the 
tunneling current responds non-linearly to the applied voltage as $I_t
\sim V_t^{2g-1}$ for non-resonant, and $I_t \sim V_t^{g-1}$ 
for resonant, tunneling \cite{Wen:1991ty,Kan13449,Wen:1993zz}. 
The temperature dependence of the tunneling conductivity is determined
by the same exponents \cite{Wen:1993zz}.\footnote{The tunneling effect
arises only when there is an assistance of the 
impurities or other interactions to absorb the other momentum along the
edge direction because electrons on two different edges have different
momentum in general. } It would be interesting to calculate the
tunneling current and  
conductivities directly in our holographic setup. This could either be
done directly by applying an 
electric field between our defects, or via the retarded correlator of
the relevant quasiparticles on the edge
\cite{Wen:1993zz}.\footnote{\label{Footnote23}According to
\cite{Wen:1993zz,Wen}, if the two edges are separated by vacuum, it is
electrons that are tunneling, and if the separation is by the FQH state,
the relevant excitations are the quasiparticles and -holes
themselves. We hence have to identify these in our model first.} There
is also a third way, employing the retarded correlator of the tunneling
operator between the edges \cite{Wen}. Of course, in
order to be consistent all these three approaches should yield the same
result. We hope to return to the calculation of the tunneling response in
the near future \cite{futurework}. 

The non-trivial correlations for the dimension five operator between 
distinct edges of the D7 branes found in \eqref{VEV616} 
exhibit a behavior that differs between 
the cases of defects separated by the YM vacuum ($k=0$) and by a 
QH state ($k\ne 0$):
correlations between insertions of the dimension 5 operator at different
edges are non-trivial (to leading order in $N$ and $\lambda$) if and
only if the two edges are connected by a D7 brane in the holographic dual. 
But the edges being connected by a D7 brane means that there is
a nontrivial YM-CS vacuum between them, while edges not
connected by any D7 brane are separated by the confining YM vacuum.
It will be interesting to 
analyze the implications of this observation for other observables
(such as \emph{e.g.} the chiral condensate) associated to defect pairs
in our model.

Tunneling experiments can also distinguish, in the AC response,
between different non-Abelian statistics  
at the same filling fraction (which in most cases determines the
Luttinger exponent $g$).\cite{Wen:1993zz}  
Another very elegant experimental setup, the two point-contact
interferometer, was proposed in \cite{Fradkin:1997ge}. In this setup,
quasi-holes can interfere along two interfering paths of a quantum
interferometer, with quasi-holes tunneling from one path to the other at
two point contacts (similar to Josephson junctions). The setup is then
equivalent to an Aharonov-Bohm type experiment, except that the
quasiholes can not only feel the quanta of magnetic flux inside the
closed loop their path is tracing, but also the non-trivial
self-statistics they have with quasiholes inserted in the loop. By
dialing the flux quanta and the number of quasiholes in the
interferometer, one can access both the effective charge and statistics
of the quasiholes. In this way, using the two  
point-contact interferometer, one can measure the VEV of closed Wilson
lines with non-Abelian statistics \cite{Fradkin:1997ge}, and ultimately
the Jones polynomial. In the holographic setup, the VEV of Wilson loops
is derived from the minimal surface of the string worldsheet ending at a
prescribed closed curve on the
boundary~\cite{Rey:1998ik,Maldacena:1998im}.\footnote{To compute such
VEV holographically, we need to specify the boundary 
condition on the minimal surface at intersecting points with
D7 branes.} It would be interesting to carry out such a calculation in
our model.  
We hope to return to this and other interesting aspects of the
model considered here
in the near future \cite{futurework}.

\section*{Acknowledgements}

We would like to thank Adi Armoni, Gerald V. Dunne, Ling-Yan Hung,
Kristan Jensen, Dmitri Kharzeev, Shiraz Minwalla,
Ioannis Papadimitriou, Shinsei Ryu, Kostas Skenderis, Tadashi
Takayanagi, Seiji Terashima, 
and Hoo-Ung Yee for helpful discussions. The work of all authors was
supported in part by the World Premier International Research Center
Initiative (WPI), MEXT, Japan. The work of S.S. was supported in part by
JSPS KAKENHI Grant Number 24540259.  
The work of R.M. was also supported in 
part by the U.S. Department of Energy under Contract 
No. DE-FG-88ER40388, as well as by the Alexander-von-Humboldt Foundation 
through a Feodor Lynen postdoctoral fellowship. 
The work of C.M.T. was supported in part by the 1000 Youth Fellowship program 
and a Fudan University start-up grant. 
M.F. is partially supported by the grants NSF-PHY-1521045 and NSF-PHY-1214341. 
We also thank the Yukawa Institute for Theoretical Physics at Kyoto
University for hospitality. Discussions during the YITP workshop
``Developments in String Theory and Quantum Field Theory''
(YITP-W-15-12) were useful to complete this work.  

\newpage

\appendix
\section{Notation}
\label{LC}

Our convention for light-cone coordinates, the Minkowski metric, the epsilon 
tensor, \emph{etc.}, are summarized as follows.
\begin{eqnarray}
 x^{\pm}=\half(x^0\pm x^1)\ ,~~
\del_\pm=\del_0\pm\del_1\ .~~
\end{eqnarray}
\begin{eqnarray}
\eta_{+-}=-2\ ,~~\eta^{+-}=-\frac{1}{2}\ ,~~
\epsilon^{01u}=-\epsilon_{01u}= +1\ ,~~
\epsilon^{+-u}=-\frac{1}{2}\ ,~~
\epsilon_{+-u}=+2\ ,
\end{eqnarray}
\begin{eqnarray}
 f_{+-}=-2f_{01}\ ,~~f_{\pm u}=f_{0u}\pm f_{1u}\ ,
\end{eqnarray}
\begin{eqnarray}
 dx^0\wedge dx^1=2 dx^-\wedge dx^+\ .
\end{eqnarray}

\medskip
We define conjugation on the product of Grassmann fields to act as
$(\xi\eta)^\dagger = \eta^\dagger\xi^\dagger$, so that, for example,
the Hermitian action for a (complex) 2d Weyl spinor $\psi_-$ is 
$S = \int d^2x\,\psi^\dagger_- \, i\p_+ \psi_- $.

\medskip
Gauge field conventions: We take the gauge field $A$ and infinitesimal
gauge parameters both to be Hermitian matrices.
The covariant derivative and field strength are given by
\be
D_\mu = \p_\mu - i A_\mu\ ,
\quad
F = dA - iA\wedge A\ ,
\ee
and gauge transformations act as 
$\delta\psi = i\alpha\psi$, $\delta A = d\alpha-i[A,\alpha]$.
When we expand in a basis for the Lie algebra, we choose an orthonormal
basis $\Tr(T^{a}T^{b})=\delta_{ab}$ (we also take our generators to be 
Hermitian), with the trace taken in the fundamental representation.

\section{Solutions of the equations of motion}\label{AppB}

\subsection{Equations of motion}
\label{eomDBI-CS}

Here we consider a single D7 brane extended along
$x^M$ ($M=0,1,u$) directions, and
the values of $y^i$ ($i=y,\tau$) are functions of $x^M$.
We are interested in the case with background metric
\begin{eqnarray}
 ds^2=G_{MN} dx^Mdx^N+G_{ij}dy^idy^j\ ,
\end{eqnarray}
where $G_{MN}$ and $G_{ij}$ are assumed to be independent of
$y^i$.
Then, the induced metric on the D7 brane is 
\begin{eqnarray}
g_{MN}=G_{MN}+G_{ij}\del_M y^i\del_N y^j\ .
\end{eqnarray}

The variation of the DBI action (\ref{DBI2}) 
under the variations of scalar fields $y^i$ and the
gauge field $a_M$ is
\begin{eqnarray}
 \delta S_{\rm DBI}
&=&
T_{\rm 3d}\int d^3 x\,\delta y^i\del_M
\left(\sqrt{-\cG}\, G_{ij}
 \cG_{\rm S}^{MN}\del_N y^j
\right)\nn\\
&&
-(2\pi\alpha')T_{\rm 3d}\int d^3 x\,\delta a_N\del_M
\left(\sqrt{-\cG}\,
\cG_{\rm A}^{MN}
\right)\nn\\
&&
-T_{\rm 3d}\int d^2 x\,\left[
\sqrt{-\cG}\,
\left( G_{ij}
 \cG_{\rm S}^{uN}\del_N y^j\,\delta y^i
-(2\pi\alpha')
 \cG_{\rm A}^{uN}\delta a_N
\right)
\right]^{z=+\infty}_{z=-\infty}\ ,
\label{delSDBI}
\end{eqnarray}
where $\cG$, $\cG_{\rm S}^{MN}$ and $\cG_{\rm A}^{MN}$ are
as defined in (\ref{cG}), (\ref{MS}) and (\ref{MA}).
The third line is the surface term for the case that there are two
boundaries at $z\ra\pm\infty$, where $z$ is defied in (\ref{z}).
The variation of the CS action (\ref{CS2}) is
\begin{eqnarray}
 \delta S_{\rm CS}
=\frac{N}{4\pi}\int d^3 x\, \epsilon^{MPN}f_{MP}\delta a_N
+\frac{N}{8\pi}\int d^2 x\,
\big[a_+\delta a_--a_-\delta a_+\big]^{z=+\infty}_{z=-\infty}\ .
\label{delSCS}
\end{eqnarray}

The equations of motion for $y^i$ and $a_N$ are
\begin{eqnarray}
\del_M\left(\sqrt{-\cG}\, G_{ij}
 \cG_{\rm S}^{MN}\del_N y^j
\right)=0\ ,
\label{geneomy}
\end{eqnarray}
and
\begin{eqnarray}
 -(2\pi\alpha')T_{\rm 3d}
\del_M\left(
\sqrt{-\cG}\,
\cG_{\rm A}^{MN}
\right)
+\frac{N}{4\pi}\epsilon^{MPN}f_{MP}=0\ .
\label{geneoma}
\end{eqnarray}
The latter equation can be written as
\begin{eqnarray}
f_{MN}=\del_M b_N-\del_N b_M\ , 
\end{eqnarray}
with
\begin{eqnarray}
 b_P\equiv \frac{\pi}{N}(2\pi\alpha')T_{\rm 3d}
\sqrt{-\cG}\,\epsilon_{MNP}\cG_{\rm A}^{MN}\ .
\end{eqnarray}
This is equivalent to the statement that
\begin{eqnarray}
a_M^{(0)}\equiv a_M-b_M
\end{eqnarray}
is a flat connection.

If we assume that $y^i$, $f_{MN}$ and all the components of
the metric only depend on $u$, the equations of motion
(\ref{geneomy}) and (\ref{geneoma}) imply
\begin{eqnarray}
\sqrt{-\cG}\, G_{ij}
\cG_{\rm S}^{uu}\del_u y^j
={\rm constant}\ ,
\label{yeom}
\end{eqnarray}
and
\begin{eqnarray}
f_{01}&=&0\ ,
\\
 -(2\pi\alpha')T_{\rm 3d}
\del_u\left(
\sqrt{-\cG}\,
\cG_{\rm A}^{u0}
\right)
+\frac{N}{2\pi}f_{1u}&=&0\ ,
\label{F1ueom}
\\
 -(2\pi\alpha')T_{\rm 3d}
\del_u\left(
\sqrt{-\cG}\,
\cG_{\rm A}^{u1}
\right)
-\frac{N}{2\pi}f_{0u}&=&0\ .
\label{F0ueom}
\end{eqnarray}

When the metric $G_{MN}$ is diagonal and
the non-zero components of the field strength are
\begin{eqnarray}
\hat e
\equiv (2\pi\alpha')f_{0u}\ ,
~~~
\hat b
\equiv (2\pi\alpha')f_{1u}\ ,
\end{eqnarray}
we have
\begin{eqnarray}
(\cG_{MN})=\left(
\begin{array}{ccc}
G_{00}&0&\hat e\\
0&G_{11}&\hat b\\
-\hat e&-\hat b &g_{uu}
\end{array}
\right)\ ,
\end{eqnarray}
\begin{eqnarray}
 \cG=G_{00}G_{11}g_{uu}+\hat b^2 G_{00}+\hat e^2 G_{11}\ ,
\label{BGcG}
\end{eqnarray}
\begin{eqnarray}
\left(\cG_{\rm S}^{MN}\right)
=\frac{1}{\cG}
\left(
\begin{array}{ccc}
\hat b^2+G_{11}g_{uu}&-\hat b\hat e&0\\
-\hat b\hat e&\hat e^2+G_{00}g_{uu}&0\\
0&0&G_{00}G_{11}
\end{array}
\right)\ ,
\end{eqnarray}
\begin{eqnarray}
\left(\cG_{\rm A}^{MN}\right)
=\frac{1}{\cG}
\left(
\begin{array}{ccc}
0&0&-\hat e G_{11}\\
0&0&-\hat b G_{00}\\
\hat e G_{11}&\hat b G_{00}&0
\end{array}
\right)\ ,
\end{eqnarray}
where
\begin{eqnarray}
 g_{uu}=G_{uu}+G_{ij}\del_u y^i\del_u y^j\ .
\label{guu}
\end{eqnarray}

In this case, the equations of motion 
(\ref{yeom}), (\ref{F1ueom}) and (\ref{F0ueom})
are
\begin{eqnarray}
c_i\equiv \frac{-G_{00}G_{11}G_{ij}\del_u y^j}{
\sqrt{-\cG}}
&=&{\rm constant}\ ,
\label{Ci}
\\
\del_u\left(\frac{\hat e G_{11}}{
\sqrt{-\cG}}
\right)
&=&-\frac{4}{R}\hat b\ ,
\label{eG11}
\\
\del_u\left(\frac{\hat b G_{00}}{
\sqrt{-\cG}}
\right)
&=&+\frac{4}{R}\hat e\ ,
\label{bG00}
\end{eqnarray}
where we have used the relation \req{T3d}.
Using \req{guu} and \req{Ci} to write $g_{uu}$ in terms of $\cG$, $G$ and 
$c_i$, we can use \req{BGcG} to conclude that
\begin{eqnarray}
\cG
=\frac{G_{00}G_{11}G_{uu}+G_{00}\hat b^2+G_{11}\hat e^2}
{1+\frac{G^{ij}c_ic_j}{G_{00}G_{11}}}\ ,
\label{detM}
\end{eqnarray}
and
\begin{eqnarray}
 \del_u y^i=G^{ij}c_j\sqrt{
\frac{G_{uu}+\frac{\hat b^2}{G_{11}}+\frac{\hat e^2}{G_{00}}}
{-G_{00}G_{11}-G^{kl}c_kc_l}}
\ .
\label{dely}
\end{eqnarray}

\subsection{Solutions for $T<T_c$.}
\label{Sol0}

For the background (\ref{metric}),
we have 
\begin{eqnarray}
 -G_{00}= G_{11}=\frac{u^2}{R^2}\ ,~~~
 G_{uu}=\frac{R^2}{u^2}\frac{1}{f(u)}\ ,~~~
 G_{yy}=\frac{u^2}{R^2}\ ,~~~
 G_{\tau\tau}=\frac{u^2}{R^2} f(u)\ .
\label{GGG2}
\end{eqnarray}

In this case,
(\ref{eG11})$\times \hat e$ +
(\ref{bG00})$\times \hat b$ implies
\begin{eqnarray}
\zeta\equiv \frac{(\hat e^2-\hat b^2)}{-\cG}\frac{u^4}{R^4}
={\rm constant}\ .
\label{wtD}
\end{eqnarray}
Then, (\ref{detM}) and (\ref{dely}) become
\begin{eqnarray}
-\cG
=\frac{u^{12}}{R^2F(u)^2}\ ,
\label{detM2}
\end{eqnarray}
and
\begin{eqnarray}
\del_u y=\frac{R^5c_y}{F(u)}\ ,~~~
\del_u\tau=
\frac{R^5c_\tau }{f(u)F(u)}
\ ,
\label{dely3}
\end{eqnarray}
where
\begin{eqnarray}
F(u)\equiv \sqrt{
u^4 f(u)\left(u^6+R^2\zeta u^4-R^6c_y^2-\frac{R^6c_\tau^2}{f(u)}
\right)}\ .
\end{eqnarray}
This function $F(u)$ agrees with (\ref{Fu}), when
\begin{eqnarray}
 \zeta=\frac{c_+c_-}{R^2}
\label{Dcc}
\end{eqnarray}
is satisfied. We will soon show that this is indeed the case.

Then, (\ref{bG00}) and (\ref{eG11}) become
\begin{eqnarray}
 \del_u\left(\frac{F(u)}{u^4}(\hat e\pm \hat b) 
\right)
&=&\mp 4(\hat e\pm \hat b)\ .
\label{deleb}
\end{eqnarray}

(\ref{dely3}) and (\ref{deleb}) can be easily integrated and we obtain
\begin{eqnarray}
y(u)=y_0+c_y\int^u_{u_{\rm min}} du'\frac{R^5}{F(u')}\ ,~~~
\tau(u)=\tau_0+c_\tau \int^u_{u_{\rm min}} du'
\frac{R^5}{f(u')F(u')}
\ ,
\end{eqnarray}
and
\begin{eqnarray}
f_{\pm u}(u)=\frac{\hat e\pm\hat b}{2\pi\alpha'}=
\frac{c_\pm}{2\pi\alpha'}\frac{u^4}{F(u)}\exp\left(
\mp 4\int_{u_{\rm min}}^u du'\frac{u'^4}{F(u')}
\right)\ ,
\label{fpmu}
\end{eqnarray}
where $y_0$, $c_y$, $\tau_0$, $c_\tau$, $c_\pm$ and $u_{\rm min}$ are
constants. With this parametrization, it is easy to check that
(\ref{Dcc}) is satisfied.
When $\del_\pm a_u=0$, (\ref{fpmu}) can be integrated as
\begin{eqnarray}
 a_\pm(u)=a_\pm^{(0)}
\pm
\frac{c_\pm}{8\pi\alpha'}
\exp\left(
\mp 4\int_{u_{\rm min}}^u du'\frac{u'^4}{F(u')}
\right)\ ,~~~
\end{eqnarray}
with constant $a_\pm^{(0)}$.

\subsection{Solutions for $T>T_c$}
\label{solT}

Here, we consider the cases with $f_{MN}=0$ and $\tau=0$.
Inserting the components
\begin{eqnarray}
 G_{00}= -\frac{u^2}{R^2}f_T(u)\ ,~~~
 G_{11}=\frac{u^2}{R^2}\ ,~~~
 G_{uu}=\frac{R^2}{u^2}\frac{1}{f_T(u)}\ ,~~~
 G_{yy}=\frac{u^2}{R^2}
\label{GGG3}
\end{eqnarray}
 of the metric (\ref{metricT}) into
(\ref{detM}) and (\ref{dely}), we obtain
\begin{eqnarray}
 -\cG=\frac{u^2}{R^2}\frac{1}{1-\frac{R^6c_y^2}{u^2(u^4-u_T^4)}}
\end{eqnarray}
and
\begin{eqnarray}
 \del_u y=\frac{R^2}{u^2}c_y\sqrt{
\frac{\frac{R^2}{u^2}\frac{1}{f_T(u)}}
{\frac{u^4}{R^4}f_T(u)-\frac{R^2}{u^2}c_y^2}
}\ .
\label{duyfinT}
\end{eqnarray}

Assuming $\del_u y=\infty$ at $u=u_*>u_T$, $c_y$ can be written as
\begin{eqnarray}
 c_y^2=\frac{u_*^2(u_*^4-u_T^4)}{R^6}\ ,
\end{eqnarray}
and (\ref{duyfinT}) becomes
\begin{eqnarray}
\del_u y=
\frac{R^2}{\sqrt{(u^4-u_T^4)
\left(\frac{u^2(u^4-u_T^4)}{u_*^2(u_*^4-u_T^4)}-1\right)}}\ .
\end{eqnarray}
Integrating this, we obtain a U-shaped solution
\begin{eqnarray} 
 y(u)=R^2\int_{u_*}^u
\frac{du'}{\sqrt{(u'^4-u_T^4)
\left(\frac{u'^2(u'^4-u_T^4)}{u_*^2(u_*^4-u_T^4)}-1\right)}}\ .
\end{eqnarray}

\end{document}